\documentclass[11pt,a4paper]{article}
\pdfoutput = 1
\usepackage{jheppub}
\usepackage{texnames}
\usepackage[T1]{fontenc}
\usepackage{hepnames}
\usepackage{siunitx}
\usepackage{color}
\usepackage{nicefrac}
\usepackage{bold-extra}

\title{Beam-beam effects on the luminosity measurement at FCC-ee}

\author[1]{Georgios Voutsinas}
\author[1]{Emmanuel Perez}
\author[2]{Mogens Dam}
\author[1]{Patrick Janot}

\affiliation[1]{CERN, EP Department, 1 Esplanade des Particules, CH-1217 Meyrin, Switzerland} 
\affiliation[2]{Niels Bohr Institute, University of Copenhagen, Blegdamsvej 17,
  2100 Copenhagen, Denmark}
  
\emailAdd{georgios.gerasimos.voutsinas@cern.ch}
\emailAdd{Emmanuel.Perez@cern.ch}
\emailAdd{dam@nbi.dk}
\emailAdd{Patrick.Janot@cern.ch}

\date{May 2019}

\begin{abstract}
{\small The first part of the physics programme of the integrated FCC (Future Circular Colliders) proposal includes measurements of Standard Model processes in $\rm e^+ e^-$ collisions (FCC-ee) with an unprecedented precision. In particular, 
the potential precision of the Z lineshape determination 
calls for a very precise measurement of the absolute luminosity, at the level of $10^{-4}$, and the precision on the relative luminosity between energy scan points around the Z pole should be an order of magnitude better.
The luminosity is principally determined from the rate of low-angle Bhabha interactions, $\rm e^+ e^- \to e^+ e^- $, where the final state electrons and positrons are detected in dedicated calorimeters covering small angles from the outgoing beam directions. Electromagnetic effects caused by the very large charge density of the beam bunches affect the effective acceptance of these luminometers in a nontrivial way. If not corrected for, these effects would lead, at the Z pole, to a systematic bias of the measured luminosity that is more than one order of magnitude larger than the desired precision. In this note, these effects are studied in detail, and methods to measure and correct for them are proposed.}
\end{abstract}

\begin{document}

\maketitle

\section{Introduction}

The FCC-ee is the first stage of a future high-energy physics programme~\cite{Benedikt:2653673} whereby particles collide in a new $100$\,km tunnel at CERN. The $\rm e^+ e^-$ collider and the experimental programme are described in the FCC-ee Conceptual Design Report (CDR)~\cite{Abada2019}. Several stages are foreseen, during which the collider is planned to run at and around the Z pole, at the WW threshold, at the ZH cross-section maximum, and at and above the $\rm t \bar{t}$ threshold. The machine delivers extremely high luminosities, in particular at the Z pole where a luminosity of $2.3 \times 10^{36}\,{\rm cm^{-2}s^{-1}}$ is expected per interaction point. To optimally exploit the very large anticipated data samples, a relative precision of $10^{-4}$ on the absolute measurement of the luminosity is desirable~\cite{Abada2019}, a factor of three better than the experimental uncertainty of the most precise measurement achieved at LEP~\cite{Abbiendi:1999zx}. Moreover, the ratio of the luminosities measured at different energy points around the Z pole must be known to within a few $10^{-5}$ (so called ``point-to-point uncertainty''). The  determination of the luminosity in $\rm e^+ e^-$ collisions usually relies on measuring the theoretically well-known rate of Bhabha interactions at small angles\footnote{The absolute theoretical uncertainty on the Bhabha cross section is expected to be reduced down to $10^{-4}$ by the time FCC-ee starts delivering collisions~\cite{Ward:2019ooj}.}, by detecting the deflected $\rm e^+$ and $\rm e^-$ in dedicated calorimeters (LumiCal) situated on each side of the interaction region. 

\begin{figure}[htb]
 \centering
\includegraphics[width=0.7\columnwidth]{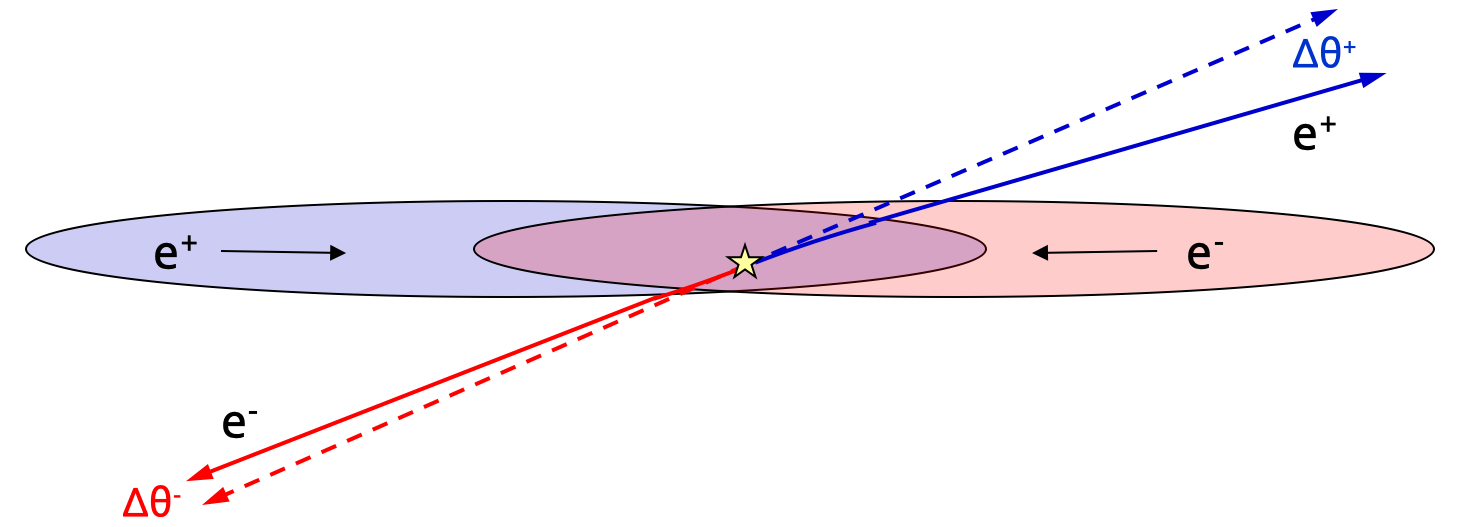} 
 \caption{Illustration of the focusing electromagnetic force that is experienced by the charged leptons emerging from a Bhabha interaction, in the case of head-on collisions. The dashed lines show the original direction of the leptons, while the full lines show their direction after the electromagnetic deflection induced by the opposite charge bunch. The case of collisions with a nonzero crossing angle is illustrated in Section~\ref{sec:bbFinalState}.
 }
 \label{fig:sketcha}
\end{figure}

As the very high FCC-ee luminosities require small beams, large electromagnetic fields are induced by the large charge density of these bunches. Electrons (positrons) in the $\rm e^-$ ($\rm e^+$) bunch feel the field from the counter-rotating $\rm e^+$ ($\rm e^-$) bunch, responsible in particular for the well-known beamstrahlung radiation.
Moreover, any charged particle present in the final state of an $\rm e^+ e^-$ interaction, if emitted at a small angle with respect to the beam direction, also feels the fields of the bunches. In particular, the final state $\rm e^+$ ($\rm e^-$) in a Bhabha interaction, emitted at a small angle from the $\rm e^+$ ($\rm e^-$) beam direction, feels an attractive force from the incoming $\rm e^-$ ($\rm e^+$) bunch, and is consequently  focused towards the beam axis\footnote{The ``repelling'' effect of the same charge beam is negligible because, in the laboratory frame, the electric and magnetic components of the Lorentz force have the same magnitude but opposite directions. In contrast, the electric and magnetic forces induced by the opposite charge beam point in the same direction and thus add up.}. This effect, illustrated in Fig.~\ref{fig:sketcha} in the case of head-on collisions, leads 
to an effective reduction of the LumiCal acceptance, as particles that would otherwise hit the detector close to its inner edge are focused to lower polar angles and, therefore, miss the detector.
As  explained below, at the Z pole, the resulting bias on the Bhabha counting rate is of the order of two per mil, a factor of $20$ larger than  the goal on the precision of the measurement. Consequently, this bias must be corrected for, and it must be known to better than $5 \%$ for this correction to contribute less than $10^{-4}$ to the systematic uncertainty on the luminosity measurement.

These ``beam-beam'' effects\footnote{Although the deflection of a final state Bhabha  $\rm e^{\pm}$, induced by the counter-rotating beam, does not correspond to the effect of one beam on the other beam, it is here still denoted as a ``beam-beam'' effect, the origin being identical to that of the genuine ``beam-beam'' interactions, which affect the initial state $\rm e^{\pm}$ in their respective bunches.} have been first studied in the context of the International Linear Collider (ILC)~\cite{Rimbault:2007zz}. The situation at FCC is, however, considerably different. In particular, FCC-ee uses the crab-waist collision scheme to achieve the expected luminosities, whereby the bunches collide with a large crossing angle ($30$\,mrad); this is in contrast to the situation at  ILC or at the Compact Linear Collider (CLIC), where the bunches are rotated by crab-crossing cavities in the vicinity of the interaction point, leading to effective head-on collisions. As shown in Section~\ref{sec:BB}, a nonzero collision crossing angle has important consequences on the beam-induced effects considered in this note.

Dedicated simulation tools, such as the {\tt{Guinea-Pig}} code~\cite{Schulte:1999tx} used here, allow these effects to be computed numerically. The correction of the bias on the luminosity could, in principle, be taken from such simulations. However, as is shown below, the correction factor depends significantly on the parameters that characterise the bunches, which may vary from bunch to bunch and during the fills, such that a numerical determination of the correction assuming averaged values for these parameters could be affected by a significant systematic uncertainty.
Moreover, since these beam-induced effects on the luminosity measurement have not been observed yet, experimental crosscheck measurements of the calculations are highly desirable. In this note, measurements are proposed to determine the luminosity correction factor with a reduced dependence on the simulation. 

This note starts with a short presentation of the experimental environment in Section~\ref{sec:ExpEnv}. In Section~\ref{sec:BB}, the beam-induced effects are described and their effects on the luminosity are explained and quantified. Section~\ref{sec:CorrectionDimuons} shows how the luminosity correction factor correlates with another observable that can be measured using the constrained kinematics of dimuon events, $\rm e^+ e^- \to \mu^+ \mu^- $. Section~\ref{sec:CorrectionInsitu} presents a complementary method to determine this correction, that relies only on $\rm e^+ e^- \to e^+ e^- $ measurements made with the luminometer. Throughout this paper, the centre-of-mass energy, $\sqrt{s}$, is taken to be 91.2\,GeV.

\section{Experimental environment}
\label{sec:ExpEnv}

The experimental environment at FCC-ee is described in detail in the CDR~\cite{Abada2019}. 
The colliding electron and positron beams cross with an angle  $\alpha = 30$\,mrad at two interaction points (IP). A detector is placed at each IP, with a solenoid that delivers a magnetic field of $2$\,T parallel to the bisector of the two beam axes, called the $z$ axis. The two beam directions define the $(x, z)$ horizontal plane. The convention used here is such that the velocity of both beams along the $x$ axis is negative, the $x$ axis pointing towards the centre of the collider ring. Two complementary central detector designs are under study. In both cases, the trajectories of charged particles are measured within a tracker down to polar angles of about $150$\,mrad with respect to the $z$ axis. The tracker is surrounded by a calorimeter and a muon detection system. The region covering polar angles below $100$\,mrad corresponds to the ``machine-detector interface'' (MDI), the design of which demands special care. A brief account  of the MDI can be found in Ref.~\cite{Boscolo:2019awb}.

The luminosity is measured from the rate of small angle Bhabha interactions, 
benefiting from the large cross-section of the Bhabha scattering process in the forward region (proportional to $ 1 / \theta^3$, where $\theta$ denotes the polar angle of the scattered $\rm e^\pm$ with respect to the outgoing $\rm e^{\pm}$ beam direction).
The Bhabha electrons are detected in a dedicated luminometer system, which consists of two calorimeters, one on each side of the interaction point. The space where these calorimeters can be installed is very tightly constrained. Indeed, in order to reach the aforementioned luminosity, the last focusing quadrupole must be very close to the IP, well within the detector volume. Moreover, as there is a $15$\,mrad angle between the momentum of the beam particles and the field of the main detector, a compensating solenoid is required in order to avoid a large blow-up of the beam emittance. In the baseline configuration~\cite{Abada2019}, the front face of this compensating solenoid is at $1.2$\,m from the IP, which basically sets the position of the end face of the LumiCal.
The LumiCal that detects $\rm e^-$ ($\rm e^+$) is centred along the direction of the outgoing $\rm e^-$ ($\rm e^+$) beam, extends along this direction between $z_{\rm{LumiCal}} = 1.074$\,m and $1.190$\,m, and covers an inner (outer) radius of $54$\,mm ($145$\,mm). For a robust energy measurement, the fiducial acceptance limits are kept away from the borders of the instrumented area, effectively reducing the acceptance to the $62 -– 88$\,mrad range.

To ensure that the luminosity measurement depends only to second order on possible misalignments and movements of the beam spot relative to the luminometer system, the method of asymmetric acceptance~\cite{Crawford:1975sw, Barbiellini} is employed. Bhabha events are selected if the $\rm e^{\pm}$ is inside a narrow acceptance in one calorimeter, and the $\rm e^{\mp}$ is inside a wide acceptance in the other. A $2$\,mrad difference between the wide and narrow acceptances is deemed adequate to accommodate possible misalignments. The narrow acceptance thus covers the angular range between $\theta_{\rm min}  = 64$\,mrad and $\theta_{\rm max} = 86$\,mrad, corresponding to a Bhabha cross section of $14$\,nb at the Z pole (compared to $40$\,nb for the Z production cross section).

The beam parameters corresponding to $\sqrt{s} = 91.2$\,GeV are given in Table~\ref{tab:parameters}, as taken from Ref.~\cite{Abada2019}. They define the nominal configuration for which the calculations presented below have been performed. Variations around this nominal configuration have also been studied. 

\begin{table}
\begin{center}
\caption{ Parameters at the Z pole that are relevant for the determination of the beam-beam effects considered here: number of particles per bunch ($N$), values of the $\beta$ function at the interaction point, in the $x$ and $y$ directions, bunch length including the bunch lengthening caused by beamstrahlung ($\sigma_s$), horizontal ($\sigma_x$) and vertical ($\sigma_y$) beam size. }
\label{tab:parameters}
\begin{tabular}{|c|c|c|c|c|c|}
    \hline
    N     &   $\beta^*_x$  & $\beta^*_y$ & $\sigma_s$  & $\sigma_x$   &   $\sigma_y$  \\
    ( $10^{10}$ )  &  ( m )  &  ( mm )  &  ( mm )  & ( $\mu$m )  &  ( nm )  \\ \hline
    17    &   0.15    &    0.8    &   12.1   &   6.36   &   28.3   \\ \hline
\end{tabular}
\end{center}
\end{table}


\section{Beam-induced effects on Bhabha events}
\label{sec:BB}

The small bunch size shown in Table~\ref{tab:parameters} leads to large charge densities, and strong electromagnetic fields are created by these bunches. Particles from a colliding bunch feel a strong force due to the field of the counter-rotating bunch, and the corresponding deflection leads to the well-known beamstrahlung radiation and ``pinch-effect''. Moreover, because of the beam crossing angle, the beam particles see their transverse momentum along the $x$ direction ($p_{x}$) increase, when they reach the interaction point. The origin of this ``kick'' and its consequences are addressed in Section~\ref{sec:bbInitialState}. In addition, charged particles emerging at small angles from an $\rm e^+ e^-$ interaction also feel the beam force, as described in Section~\ref{sec:bbFinalState}. Section~\ref{sec:tools} describes the tools that have been used to compute these effects.


\subsection{Numerical calculations}
\label{sec:tools}

\subsubsection{The {\texttt{\textbf{Guinea-Pig}}} simulation program}
\label{sec:GuineaPig}

The {\tt Guinea-Pig} code~\cite{Schulte:1999tx} was initially developed in the mid-nineties to simulate the beam-beam effects and the beam background production in the interaction region of future electron-positron colliders. It has been used extensively  since then. {\tt Guinea-Pig} groups particles from the incoming bunches into macro-particles, slices each beam longitudinally, and divides the transverse plane into cells by a ``grid''. The macro-particles are initially distributed over the slices and the grid, and are tracked through the collision, the fields being computed at the grid points at each step of this tracking. Because of the crossing angle, this grid has to be quite large in order to encompass the $2$ to $3 \sigma_{s}$ envelope of the beam, and a size in $x$ of $d_x = 150 \sigma_x  > 2 \times 2 \sigma_s / (\alpha/2)$ has been chosen. The grid dimension along the $y$ direction should account for the very small $\beta^*_y$, and a size of $d_y = ( 2 \times 2 \sigma_s / \beta^*_y ) \times \sigma_y = 60 \sigma_y$ is used here.  
The number of cells are such that the cell size, in both the $x$ and $y$ dimensions, amounts to about $10 \%$ of the transverse beam size at the interaction point. 

In the context of the studies reported in Ref.~\cite{Rimbault:2007zz}, the C++ version of {\tt Guinea-Pig} has been extended in order to track Bhabha events, provided by external generators like {\tt BHWIDE}~\cite{Jadach:1995nk}, in the field of the colliding bunches. This version of {\tt Guinea-Pig} is used here. An input Bhabha event is associated to one of the $\rm e^+ \rm e^-$ interactions, i.e.\ is assigned a spatial vertex and an interaction time according to their probability density. Beamstrahlung, that causes the energy of the initial state particles to be reduced, as well as the electromagnetic deflection of these particles due to the field of the opposite bunch, are taken into account by rescaling, boosting and rotating the generated Bhabha event~\cite{Rimbault:2007zz}. The electron and positron that come out from this Bhabha interaction see their four-momenta corrected when these transformations are applied. They are subsequently transported as they move forward: the final state $\rm e^-$ ($\rm e^+$) potentially crosses a significant part of the $\rm e^+$ ($\rm e^-$) bunch, or travels for some time in the vicinity of this bunch and, thereby, feels a deflection force. 

Since the particles of a given slice $i$ of the $\rm e^-$ bunch  feel the field created by each slice $j$ of the $\rm e^+$ beam, in turn, as the bunches move along, the execution time of the program scales with the number of $(i, j)$ combinations, i.e.\ it scales quadratically with the number of slices. With the parameters given in Table~\ref{tab:parameters}, 
the position of the interaction vertex along the $z$ direction ($z_{\rm{vtx}}$) follows a Gaussian distribution whose standard deviation, given by
\begin{equation}
    \Sigma_{z} = \left[ 2 \left( \frac{ \sin^2 \alpha/2}{\sigma^2_{x}} + \frac{\cos^2 \alpha/2}{ \sigma^2_{s} } \right)\right]^{-1/2} ,
  \label{eq:SigmaIR}  
\end{equation}
amounts to $0.3$\,mm only. 
When the $ 2 \sigma_s$ envelope of the beam is considered, at least $160$ slices are thus needed to ensure that the size of each slice is smaller than $\Sigma_{z}$.

Since the $\rm e^{\pm}$ that emerge from a Bhabha interaction are emitted with a non vanishing, albeit small, polar angle, they may exit the grid mentioned above, designed to contain the beams  and in which the fields are computed, before the tracking ends. For this reason, the program can also extend the calculations of the fields to ``extra'' grids.
In the version of the code used here, up to six extra grids can be defined, which cover a larger and larger spatial volume, with a decreasing granularity. 
These  extra grids can be designed such that the largest one safely contains the trajectory of Bhabha electrons during the whole tracking time (e.g.\ up to a maximal time $t_{\rm{max}} = 3 \sigma_{s} / c$, the time origin being given by the time when the centres of the two bunches overlap).
However, in the general case, if a charged particle exits the largest grid considered by the program before the tracking ends, values for the fields are still determined, the beams being then approximated by line charges. The execution time of the program scales linearly with the number of grids, since the calculation of the fields at each point of these grids is the most time consuming operation.

With a total of seven grids and 300 slices, the field calculations take about one week on a $2.6$\,GHz {\tt Intel i$7$} processor. Running with one single grid and $750$  slices takes as long. Unless explicitly stated otherwise, the {\tt Guinea-Pig} simulations shown below were obtained with the latter setting.
We made this choice since, as shown below, the approximation of using only one grid remains accurate for the angular range considered here.

\subsubsection{Analytic determination of the average effects}
\label{sec:NumericalCalc}

A numerical integration code has also been developed, that uses the Bassetti-Erskine formulae~\cite{Bassetti:1980by} for the field created by a Gaussian bunch to determine the average effects that a particle would feel. The formalism is described in Ref.~\cite{Keil:1994dk}. The particle is defined by its velocity and spatial coordinates at a given time $t_0$. The momentum kick that it gets between $t_0$ and a later time is obtained by integrating the Lorentz force that it feels during this time interval. The {\tt CUBA} library~\cite{Hahn:2004fe} is used to perform the numerical integrations. For the calculation of the Faddeeva function, $w(z) = \exp( -z^2 ) ( 1 - {\rm{erf}}( -i z ) )$, which  enters the expression of the electromagnetic field created by a two-dimensional Gaussian bunch, the implementation provided by the RooFit package~\cite{Karbach:2014qba} has been used.


\subsection{Effects on the initial state particles}
\label{sec:bbInitialState}

The effect of beamstrahlung on the energy of the interacting particles has been studied in detail in Ref.~\cite{blondel2019polarization}. For the $\rm Z$ pole running parameters, the average energy loss of the $45.6$\,GeV beam particles amounts to $310$\,keV only. 
Consequently, the situation at FCC is very different from what would happen at ILC where large and asymmetric radiations off the incoming $\rm e^-$ and $\rm e^+$ legs would lead to a longitudinal boost of the $\rm e^+ \rm e^-$ centre-of-mass frame and to a large acollinearity of the final state~\cite{Rimbault:2007zz}. At FCC, the very small reduction of the $\rm e^{\pm}$ energies due to beamstrahlung radiation has a negligible impact on the fraction of Bhabha electrons that emerges within the acceptance of the LumiCal.

\begin{figure}[htbp]
\begin{center}
\includegraphics[width=0.7\columnwidth]{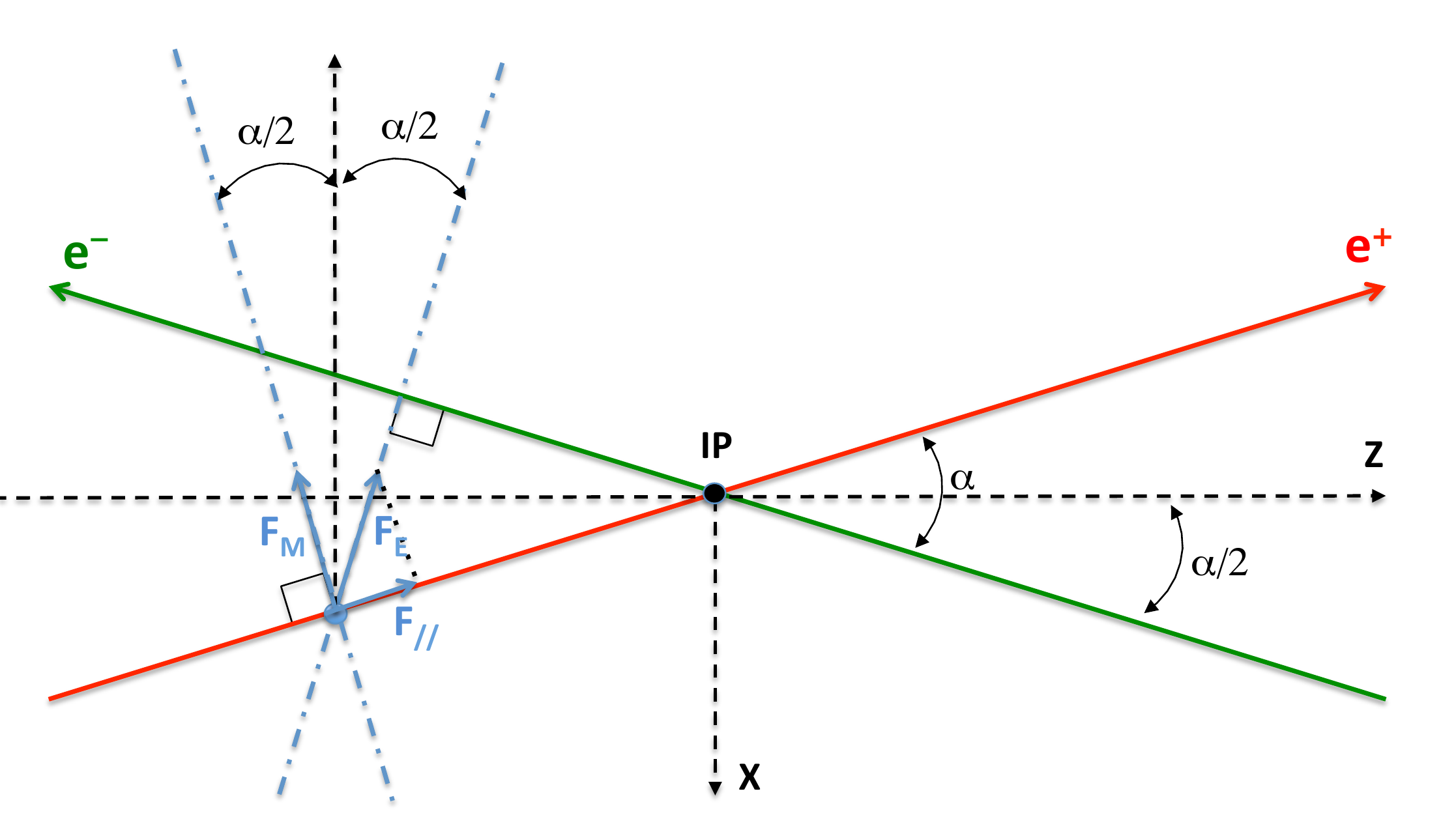}
\caption{\small Schematic view of the electric and magnetic attractive Lorentz forces $\vec{F}_E$ and $\vec{F}_M$ acting on each positron from the opposite electron bunch, upon bunch crossing at the interaction point (IP). Similar forces from the positron bunch affect each electron. From Ref.~\cite{blondel2019polarization}.  
}
\label{fig:energykick-schema}
\end{center}
\end{figure}

Another important effect, also detailed in Ref.~\cite{blondel2019polarization}, is however taking place due primarily to the crossing angle. Figure~\ref{fig:energykick-schema} illustrates the force experienced by an $\rm e^+$ in the positron bunch, due to the fields created by the counter-rotating electron bunch. The electrons being ultra-relativistic, the fields that they induce are compressed into a plane, perpendicular to their trajectory. Consequently, the electric component $\vec{F}_E$ of the Lorentz force experienced by the positron is orthogonal to the $\rm e^-$ direction. The magnetic component of the force, $\vec{F}_M$, is on the other hand perpendicular to the $\rm e^+$ trajectory. In the laboratory frame, the magnetic field created by the $\beta = 1$ electrons is $ B = E/c $, such that $F_E = F_M$. The resulting vector sum is a force that is parallel to the $x$ axis, which accelerates the positron before it reaches the IP ($F_x < 0$, as illustrated in Fig.~\ref{fig:energykick-schema}), and decelerates it after it has crossed the IP ($F_x > 0$). When integrated over a large interval around the time $t=0$ when the centres of the two bunches overlap,
and averaged over all positrons in the bunch, the resulting momentum kick $k_x$ vanishes. However, at the time when the particles interact, only negative components $k_x < 0$ have been integrated. This truncated integral results in a boost of the $\rm e^+ \rm e^-$ system 
in the horizontal direction (along $-x$),
in addition to that induced by the nominal crossing angle -- or, equivalently, to an effective increase of the crossing angle.

\begin{figure}[htbp]
\begin{center}
\begin{tabular}{cc}
\includegraphics[width=0.48\columnwidth]{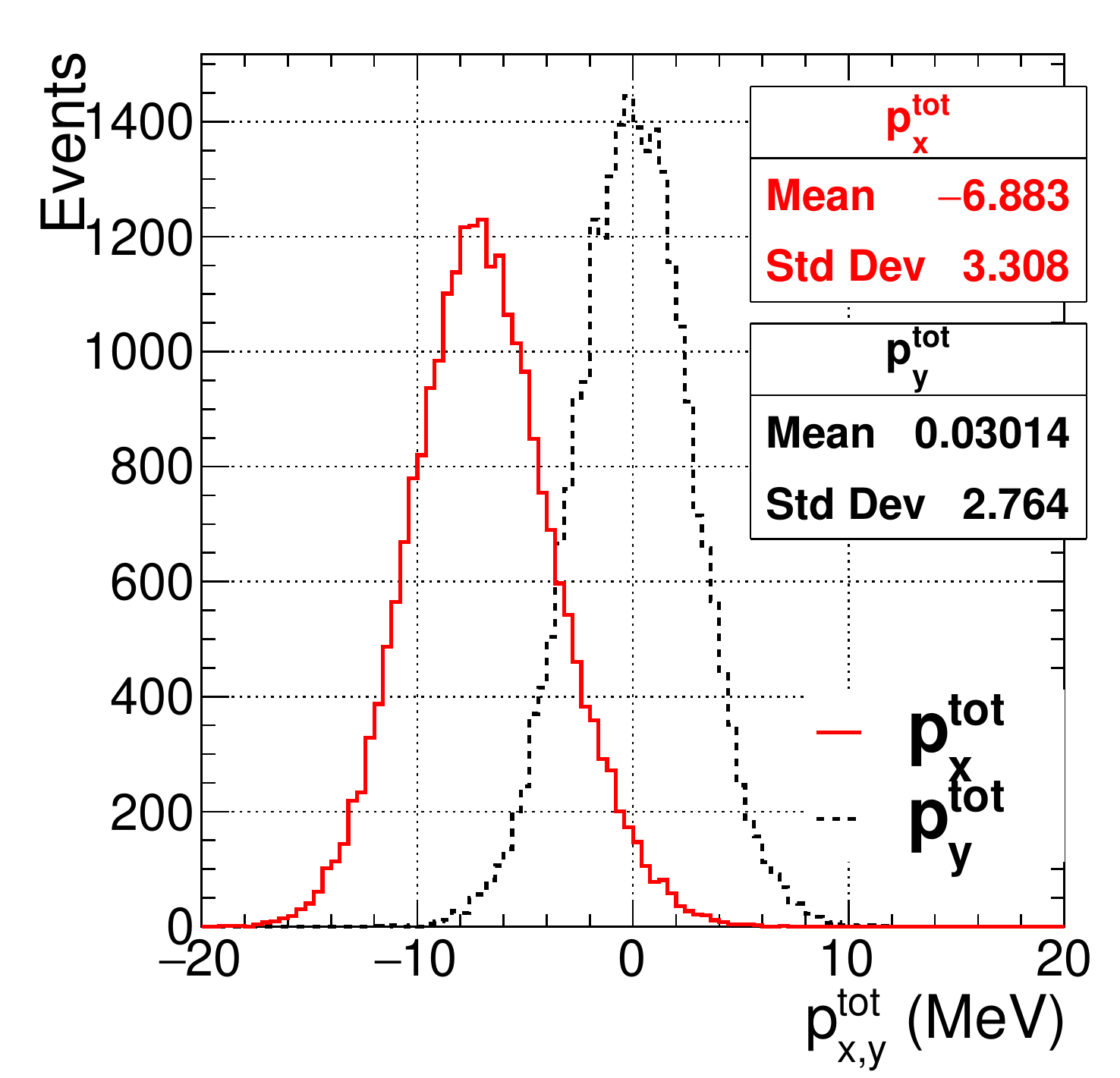} &
\includegraphics[width=0.48\columnwidth]{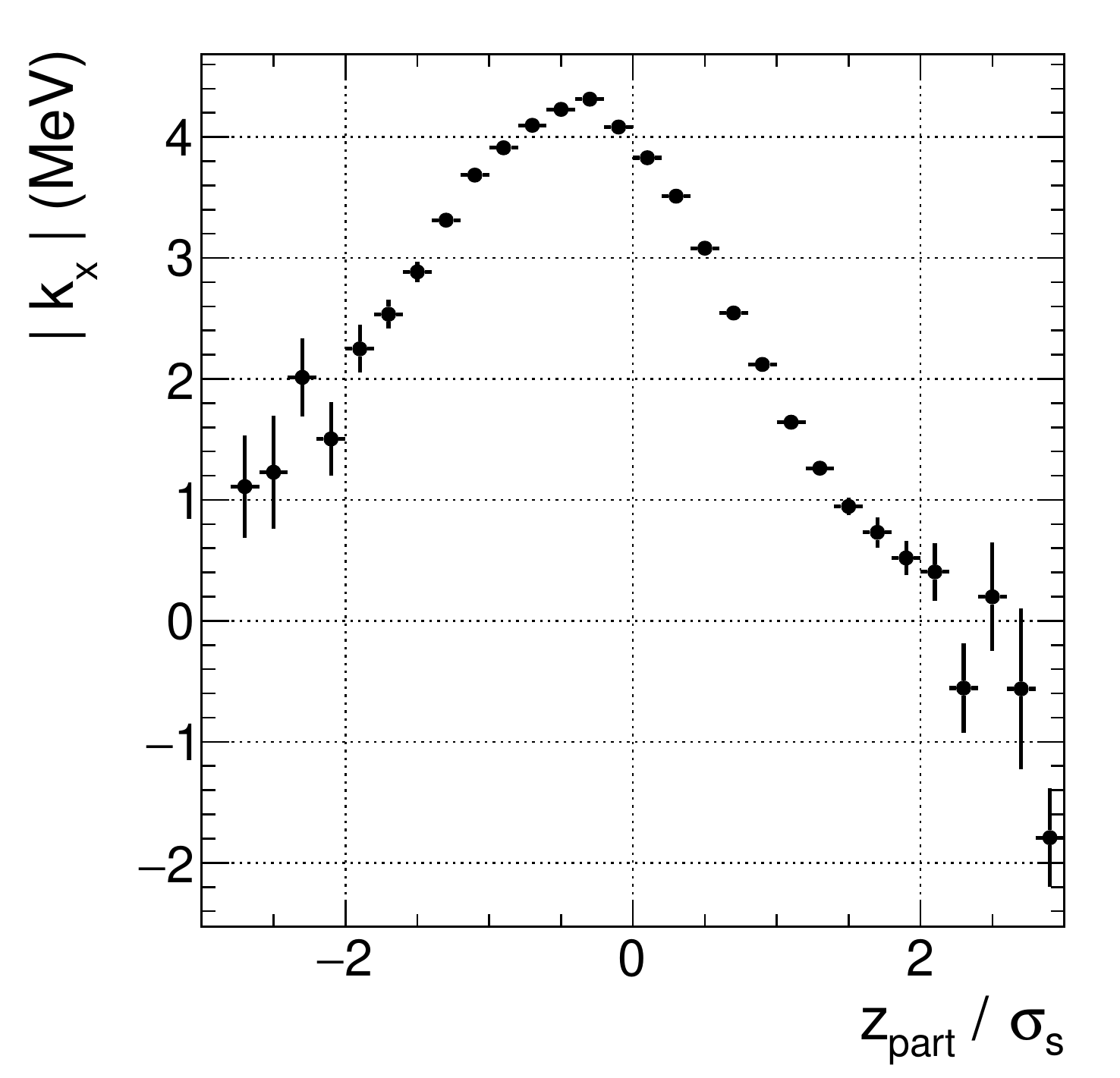}
\end{tabular}
\caption{\small Left: Distribution of the transverse components of the total momentum of $\rm e^+ \rm e^-$ events, as predicted by {\tt Guinea-Pig}. The direction of the $x$ axis is chosen such that the bunches move with a negative velocity along $x$, and the $p_x$ component is given in a frame that moves along $-x$ with a velocity $v = c \times \sin \alpha/2 $.   Right: Kick along $-x$ that is acquired by an incoming $\rm e^-$ (or $\rm e^+$) as a function of its longitudinal position $z_{\rm{part}}$ within the bunch, in units of the bunch length. Positive (negative) values of $z_{\rm{part}}$ correspond to the head (tail) of the bunch.
}
\label{fig:pxkick-GP}%
\end{center}
\end{figure}

This boost is illustrated in the left panel of Fig.~\ref{fig:pxkick-GP}, which shows the distribution of the transverse components of the total momentum of $\rm e^+ \rm e^-$ events as predicted by {\tt Guinea-Pig}. These distributions were obtained from Bhabha events, but would be the same for any other final state. The horizontal component $p_{x}^{\rm{tot}}$ is given in a frame that moves with a velocity $v = c \times \sin \alpha/2 $ along $-x$, i.e.\ a frame in which, in the absence of beam-beam effects, the bunches would have no transverse momentum.
While the mean of the distribution of $p_{y}^{\rm{tot}}$ is consistent with zero within the statistical uncertainties, the average of $p_{x}^{\rm{tot}}$ is shifted by about $7$\,MeV. This shift corresponds to a kick $ \mid k_x \mid = 3.5$\,MeV acquired by both the $\rm e^-$ and the $\rm e^+$ by the time they interact. As a comparison, the momentum along  $-x$  of the incoming particles due to the nominal crossing angle is equal to $E_{\rm{beam}} \times \sin \alpha /2 \simeq 700$\,MeV, where $E_{\rm{beam}} = \sqrt{s}/2$. The right panel of  Fig.~\ref{fig:pxkick-GP} shows how the kick acquired by an incoming $\rm e^+$ or $\rm e^-$ varies with the longitudinal position of the particle within the bunch. As expected, the kick is smaller for particles in the head or in the tail of a bunch, than for those that are in the middle of it.

\begin{figure}[htbp]
\begin{center}
\begin{tabular}{ccc}
\includegraphics[width=0.315\columnwidth]{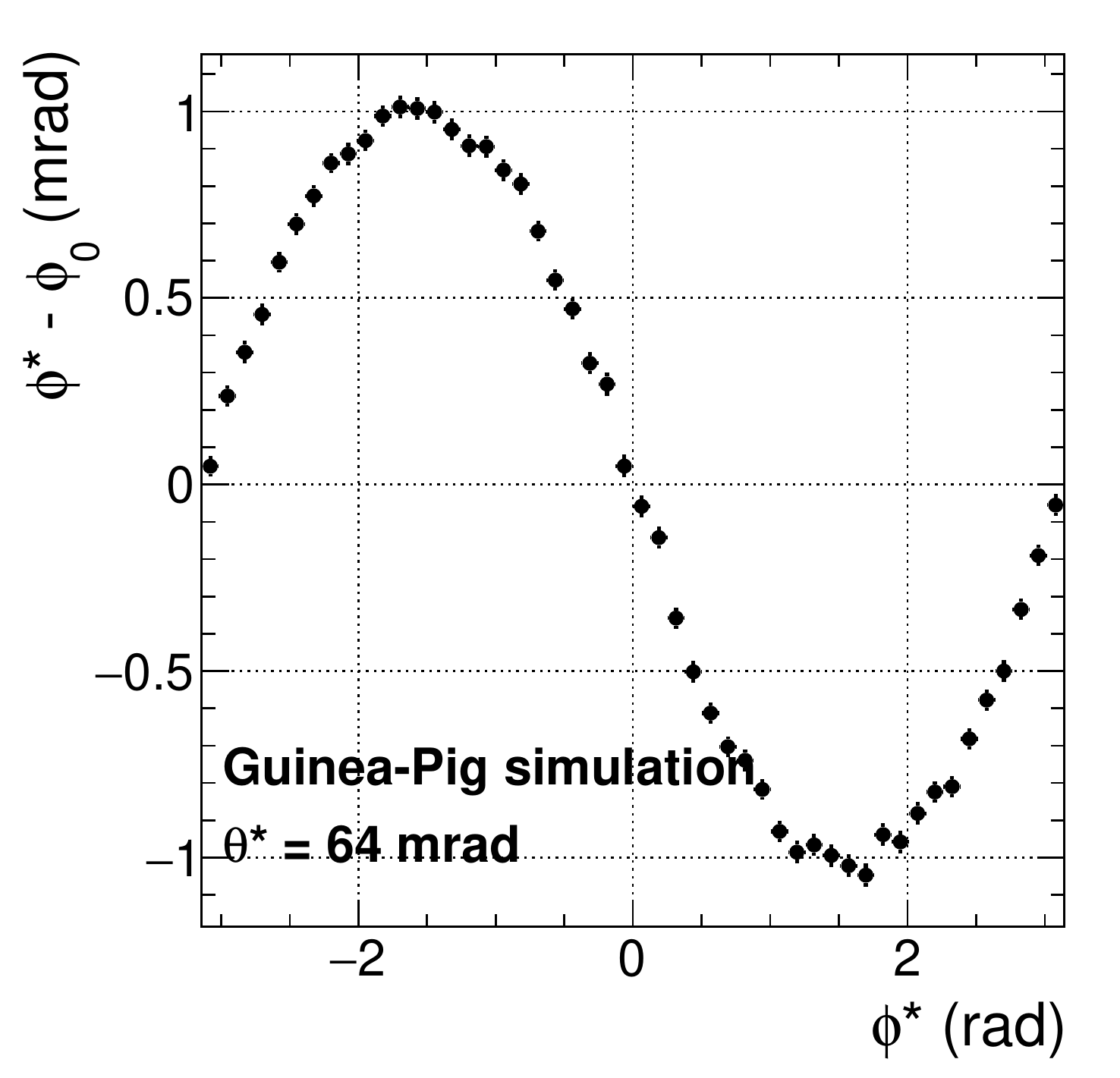} &
\includegraphics[width=0.315\columnwidth]{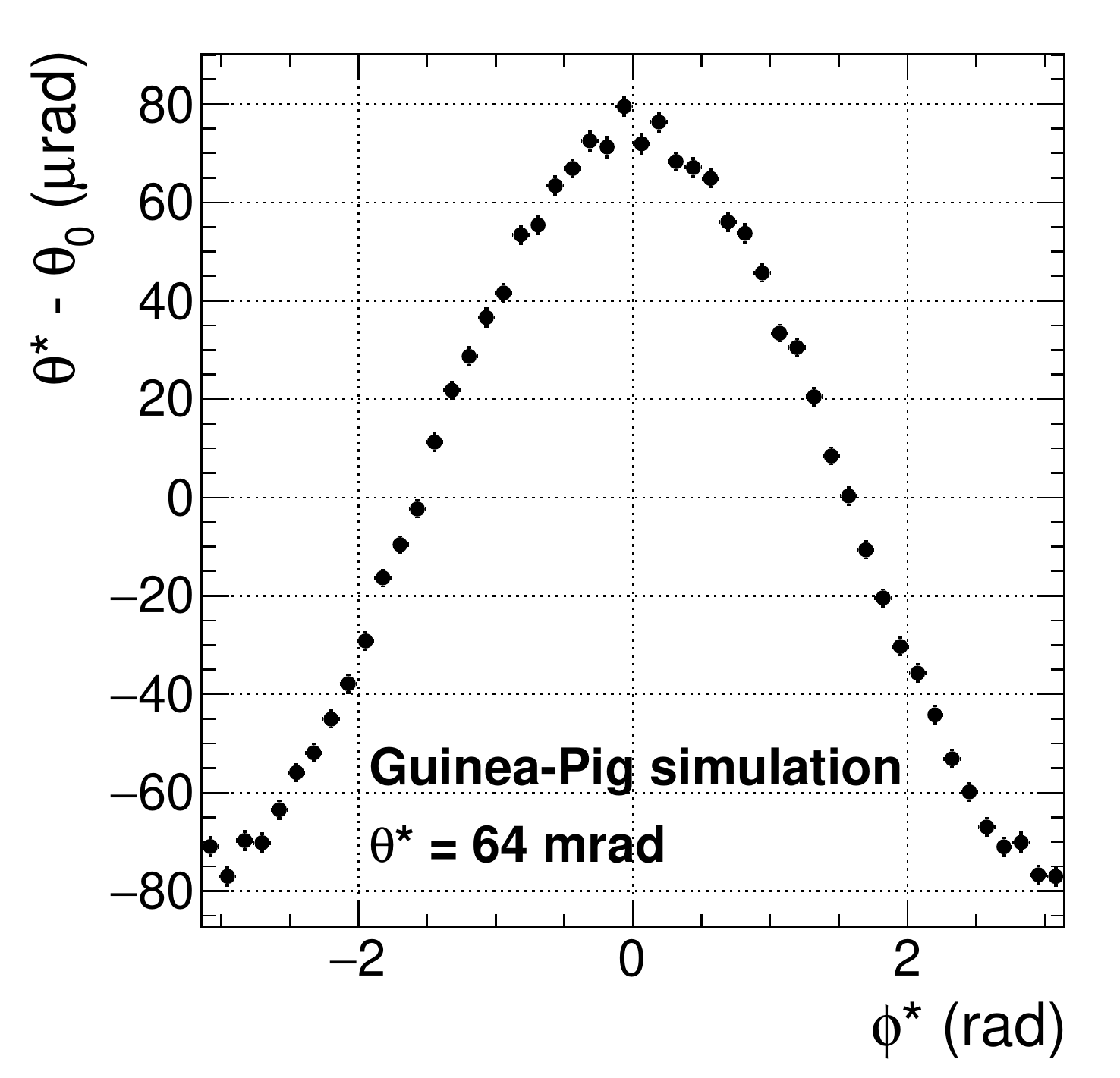} &
\includegraphics[width=0.315\columnwidth]{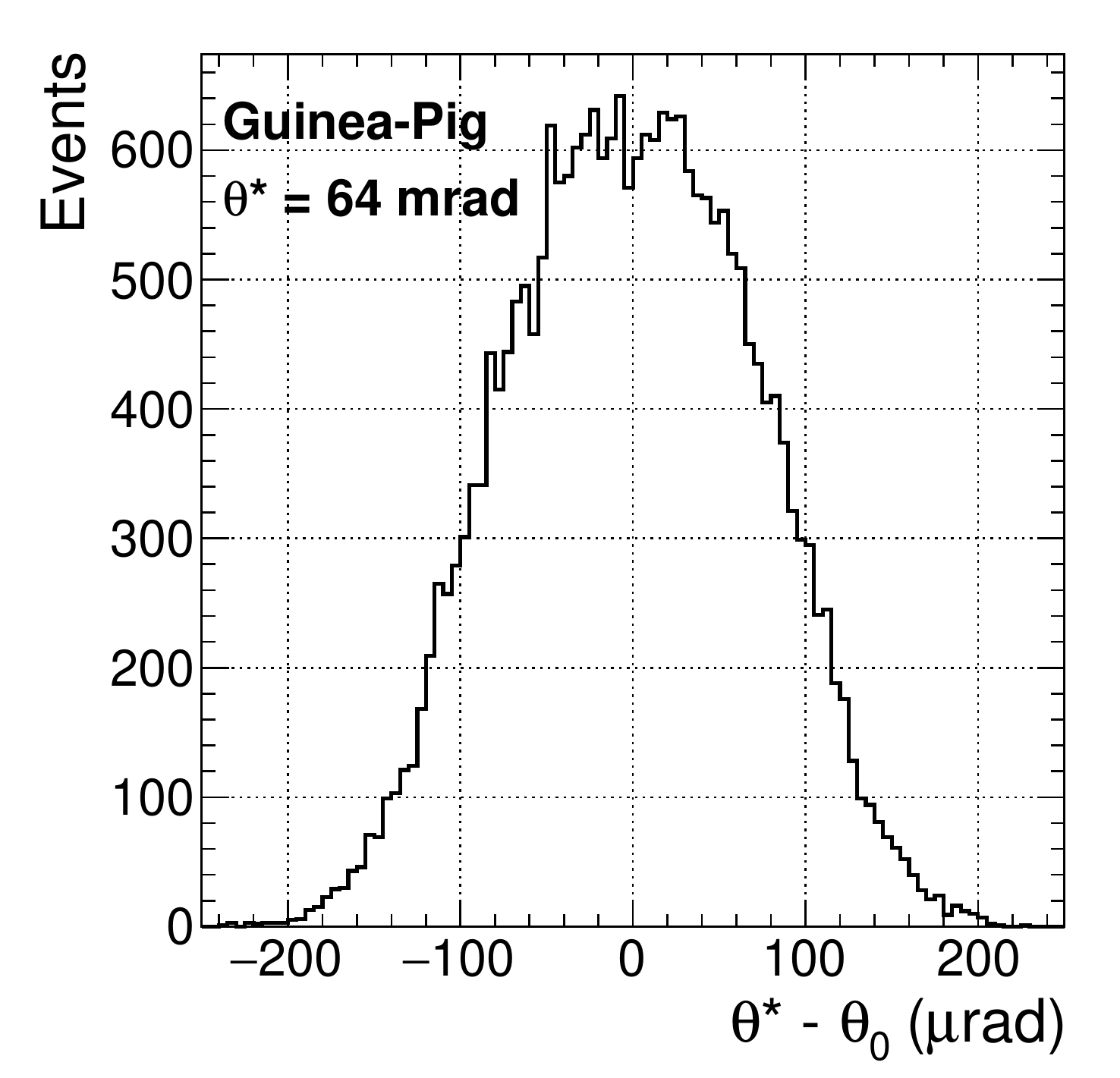} 
\end{tabular}
\caption{\small Left: Average of the difference between the $\rm e^-$ azimuthal angle after the beam-beam effects in the initial state ($\phi_{0}$) and the angle at the generator level ($\phi^{*}$), as a function of the latter, for $45.6$\,GeV electrons from leading-order Bhabha interactions, emitted at $64$\,mrad 
with respect to the electron beam direction.
Middle: idem but for the $\rm e^-$ polar angle. Right: Distribution of the shift in polar angle induced by the beam-beam effects in the initial state.
}
\label{fig:deltaTheta_and_phi_kick-GP}%
\end{center}
\end{figure}

This $p_x$ kick leads to a modification of the kinematics of the particles that emerge from the Bhabha interaction. When the $p_x$ of a final state  $\rm e^{\pm}$ is shifted by an amount $\delta p_x$, its polar angle $\theta$ (defined with respect to the direction of the $\rm e^{\pm}$ beam) and its azimuthal angle $\phi$ are shifted according to :
$$
\theta_0 = \theta^* + \frac{ \delta p_x} {\mid p_z \mid } \cos \phi  \qquad , \qquad
\phi_0 =  \phi^* + \frac{ \delta p_x} {p_T} \sin \phi
$$
with $ \left<  \delta p_x \right > = k_x$ for the $\rm e^{\pm}$ emerging from a leading-order Bhabha interaction, and where $p_T$ denotes the $\rm e^{\pm}$ transverse momentum with respect to the direction of the $\rm e^\pm$ beam. The $^*$ superscripts denote the angles prior to this boost, 
while the nought subscripts label the kinematic quantities of the particles as they emerge from the interaction. The {\tt Guinea-Pig} tracking of $45.6$\,GeV electrons 
emitted at an angle of $64$\,mrad with respect to the electron beam direction
agrees with these formulae, as shown in Fig.~\ref{fig:deltaTheta_and_phi_kick-GP}.
The angular shifts of the outcoming $\rm e^-$ and $\rm e^+$ go in the opposite direction both in $\phi$ and in $\theta$ (i.e., if the kick increases the angle of the $\rm e^-$ with respect to the outgoing $\rm e^-$ beam, the $\rm e^+$ is focused closer to the outgoing $\rm e^+$ beam direction).
When  averaged over the azimuthal angle of the $\rm e^{\pm}$, the $p_x$-kick smears the initial $\theta$ distribution but does not bias its mean value, as shown in the right panel of Fig.~\ref{fig:deltaTheta_and_phi_kick-GP}. 
These effects are similar to those of a misalignment of the luminometer system with respect to the IP along the $x$ direction. The kick expected for the nominal running parameters at the $\rm Z$ pole is equivalent to a misalignment of $$ \delta_x = \frac{ k_x} {E_{\rm{beam}}} \cdot z_{\rm{LumiCal}} \sim 80\,\mu{\rm m}.$$ With the method of asymmetric acceptance mentioned in Section~\ref{sec:ExpEnv}, the resulting relative bias on the luminosity depends only quadratically on $ \delta_x$ or $k_x$, and the $p_x$-kick induced by the beam-beam effects in the initial state has a negligible effect (a few $10^{-6}$) on the   measurement.

\subsection{Effects on the final state particles}
\label{sec:bbFinalState}

\subsubsection{Characterisation of the effect}

\begin{figure}[htbp]
\begin{center}
\begin{tabular}{ccc}
\includegraphics[width=0.315\columnwidth]{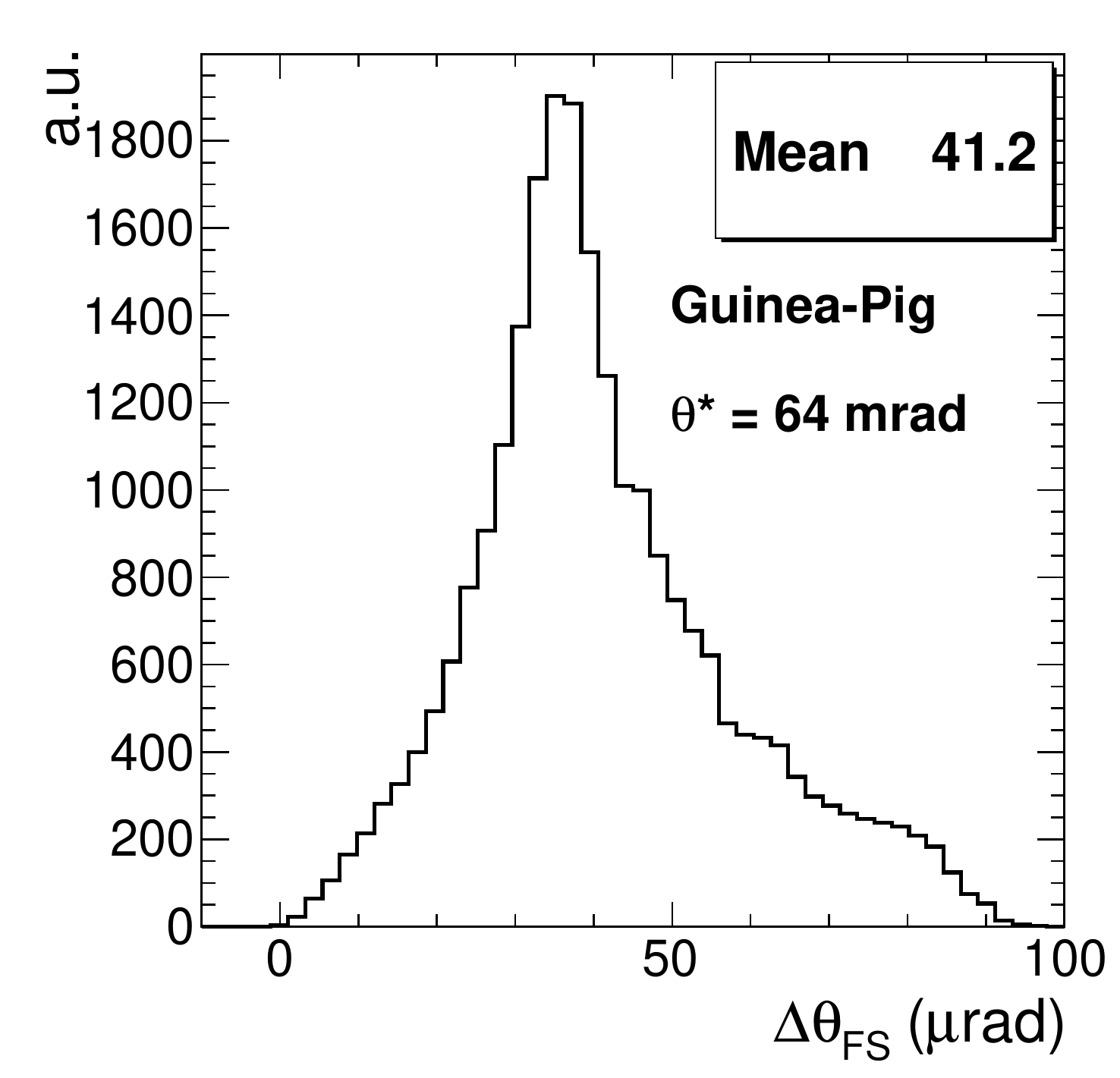} &
\includegraphics[width=0.315\columnwidth]{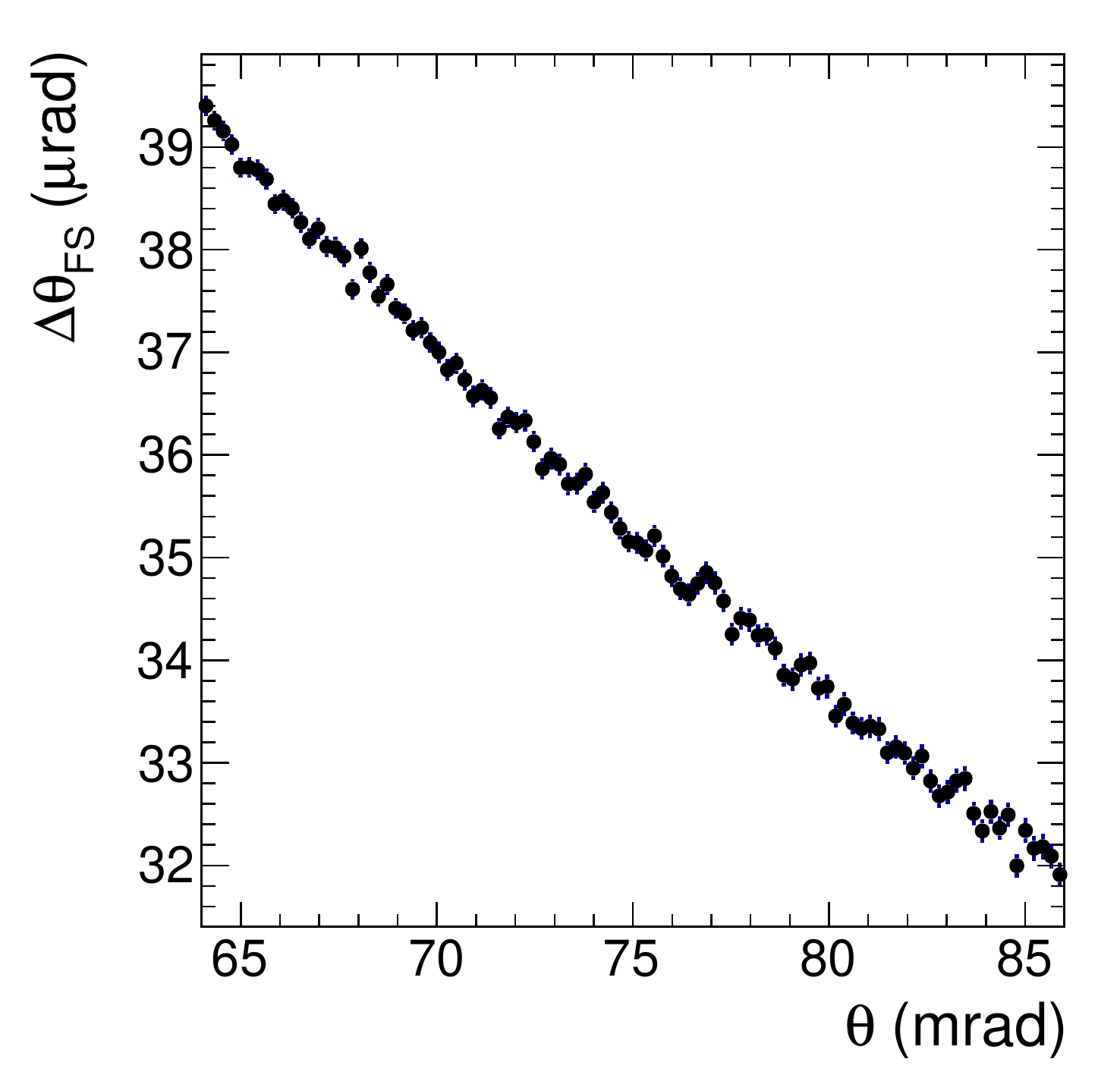} &
\includegraphics[width=0.315\columnwidth]{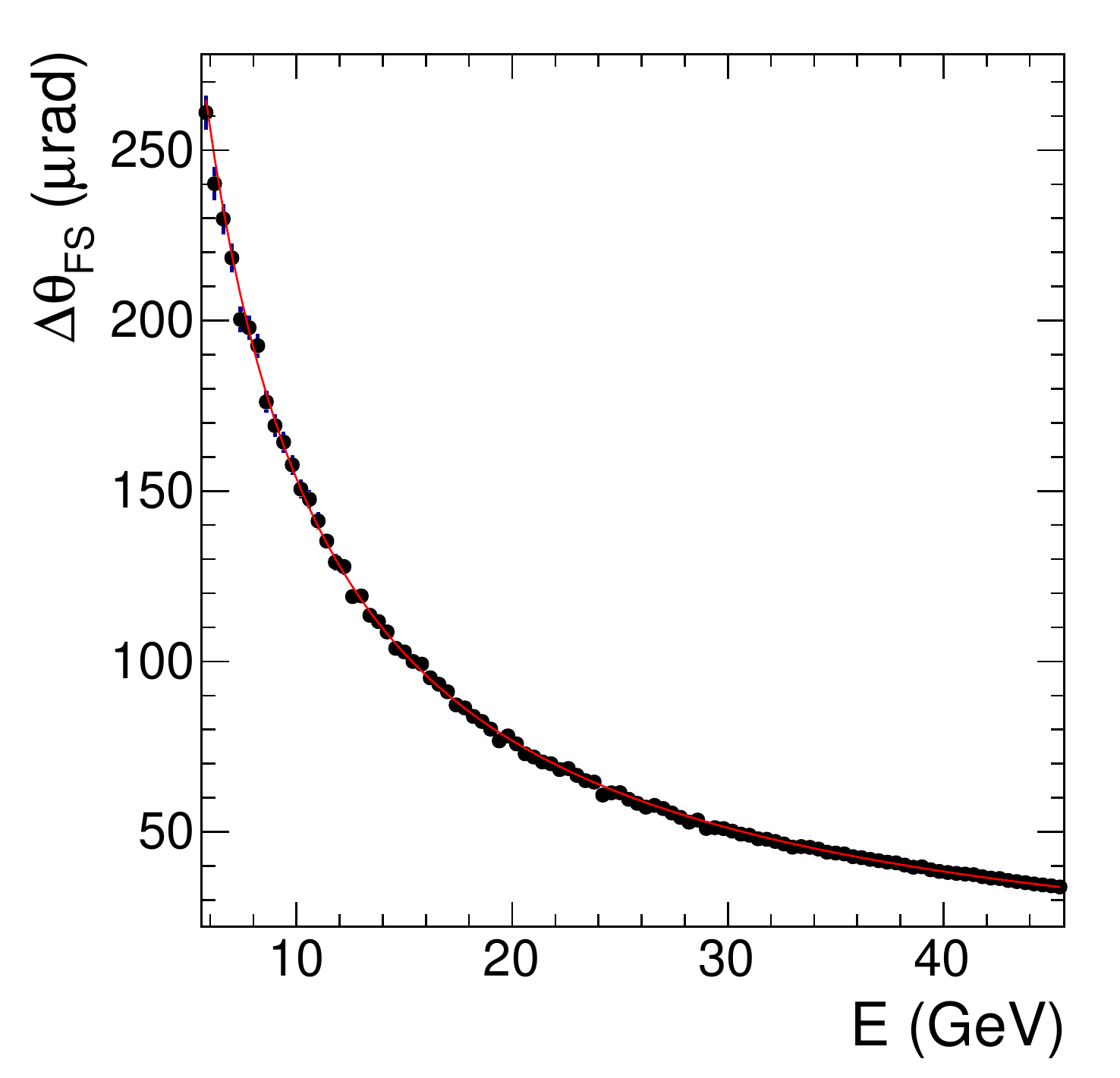}
\end{tabular}
\caption{\small Left: Distribution of the angular focusing $\Delta \theta_{\rm{FS}}$ for electrons with $E = 45.6$\,GeV and emitted at a fixed angle $\theta^* = 64$\,mrad with respect to the electron beam direction. Middle and right : $\Delta \theta_{\rm{FS}}$ as a function of (middle) the polar angle and (left) the energy of the outcoming lepton in Bhabha events. The events were generated with the {\tt BHWIDE} program, in the angular range $64\,{\rm mrad} < \theta^* < 86\,{\rm mrad}$. A fit to $\Delta \theta_{\rm{FS}} \propto 1/E$ is overlaid on the right plot.
}
\label{fig:deltaTheta_BHWIDE}
\end{center}
\end{figure}

The field of the opposite charge bunch deflects also the electrons and positrons emerging from a Bhabha interaction.
The left panel of Fig.~\ref{fig:deltaTheta_BHWIDE} shows the distribution of the angular deflection $\Delta \theta_{\rm{FS}}$ of $45.6$\.GeV electrons 
emitted at
a fixed angle of $64$\,mrad with respect to the electron beam direction, as predicted by {\tt Guinea-Pig}. It is defined as the difference between the polar angle of the outgoing electron before and after this deflection, $\Delta \theta_{\rm{FS}} = \theta_{0} - \theta$ where $\theta$ denotes the final angle, such that a positive quantity corresponds to a focusing deflection along the beam direction. For electrons emerging close to the lower (upper) edge of the fiducial LumiCal acceptance, $\theta \simeq \theta_{\rm{min}}$ ($\theta \simeq \theta_{\rm{max}}$), the average deflection amounts to $41.2\,\mu$rad ($34.8\,\mu$rad). The net effect is that the number of electrons detected in the LumiCal, in the range  $\theta_{\rm{min}} < \theta < \theta_{\rm{max}}$, is smaller than the number of Bhabha electrons emitted within this range, which leads to an underestimation of the luminosity. From the expression of the counting rate in the LumiCal: $$ N \propto \int_{ \theta_{\rm min}}^{\theta_{\rm max}} \frac{ d \theta} { \theta^3}  ,$$ the bias induced by this angular deflection reads:
\begin{equation}
 \Delta N / N = \frac{ -2 } { \theta_{\rm{min}}^{-2} - \theta_{\rm{max}}^{-2} } \left( \frac{ \Delta \theta_{\rm{FS}} (\theta = \theta_{\rm{min}})}{\theta_{\rm{min}}^3} - \frac{ \Delta \theta_{\rm{FS}} (\theta = \theta_{\rm{max}})}{\theta_{\rm{max}}^3} \right )  ,
 \label{eq:lumiBias}
 \end{equation}
which, numerically, leads to a bias on the measured luminosity of $\Delta L / L \simeq - 0.19 \%$, almost $20$ times larger than the target precision on the luminosity measurement. This bias must therefore be corrected for, and the correction factor should be known with a relative uncertainty of less than $5 \%$ to ensure a residual systematic uncertainty smaller than $10^{-4}$ on the measured luminosity.

The middle and right panels in Fig.~\ref{fig:deltaTheta_BHWIDE} show the angular and energy dependence of this deflection $\Delta \theta_{\rm{FS}}$, as seen in  Bhabha events generated with the {\tt BHWIDE} program within the phase space of the measurement. As expected, the deflection gets smaller when the polar angle of the electrons increases, and it increases as $1/E$ when their energy $E$ decreases.

\begin{figure}[bhtp]
\begin{center}
\begin{tabular}{cc}
\includegraphics[width=0.5\columnwidth]{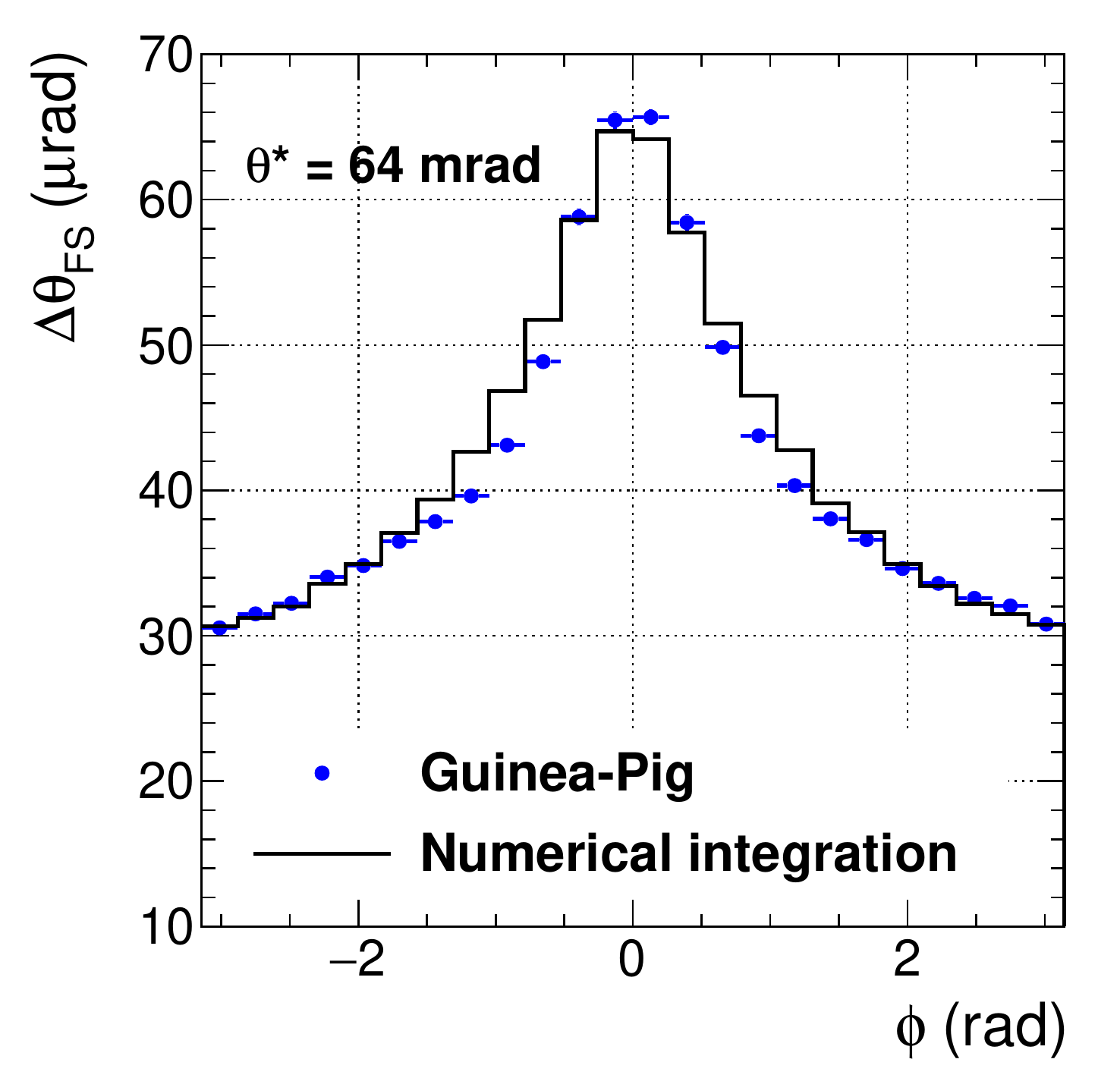} &
\includegraphics[width=0.5\columnwidth]{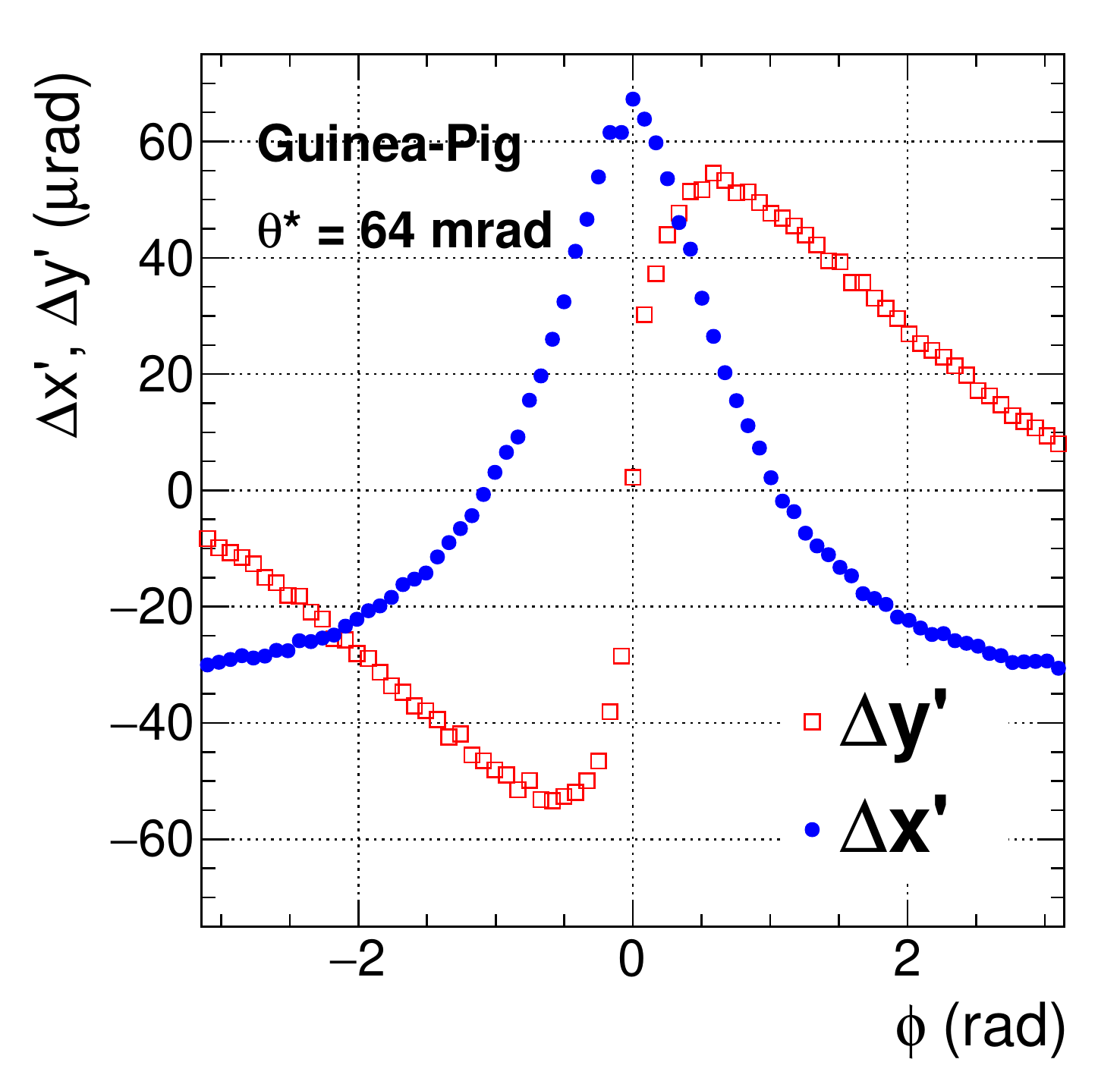}
\end{tabular}
\caption{\small Left: $\Delta \theta_{\rm{FS}}$ for $45.6$\,GeV electrons produced at $\theta^* = 64$\,mrad as a function of their azimuthal angle, as predicted by {\tt Guinea-Pig} 
and by a numerical integration of the average Lorentz force felt by the electrons. Right: Angular deflections along the $x$ and $y$ directions, under the same conditions. 
}
\label{fig:deltaTheta_vs_phi_GP_numerical}
\end{center}
\end{figure}

\begin{figure}[htb]
 \centering
 \includegraphics[width=0.9\textwidth,keepaspectratio]{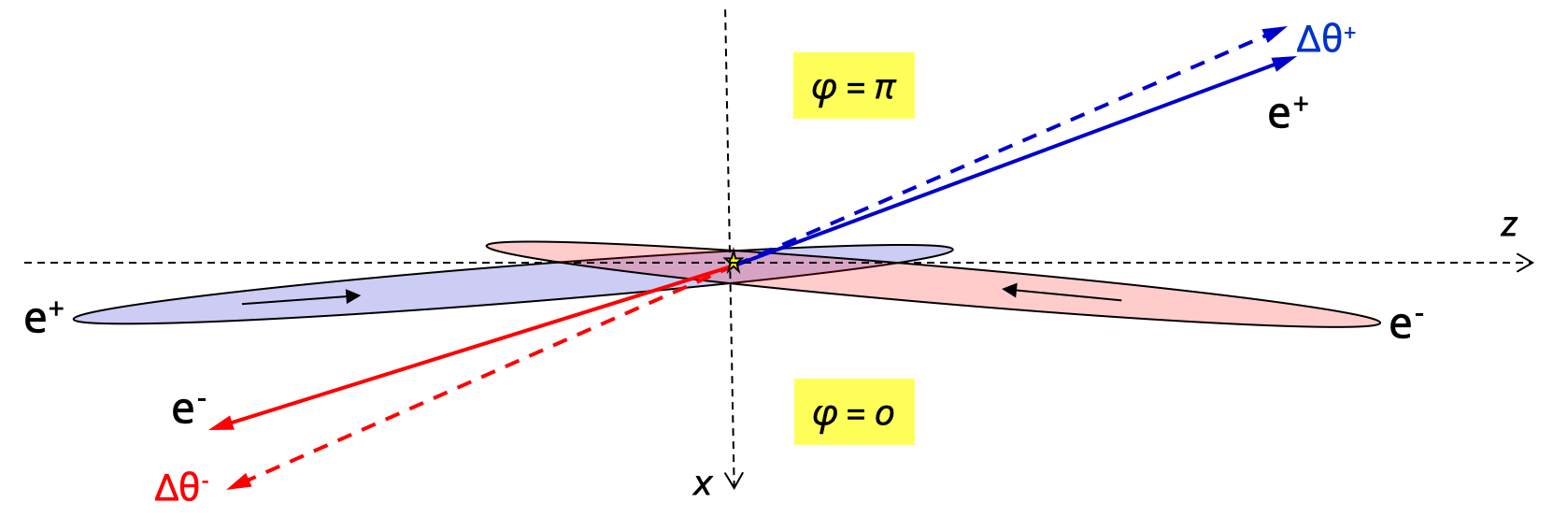}
 \caption{Electrons (positrons) emitted along the positive $x$ direction are
   closer to the opposite charge positron (electron) bunch, and therefore their
   focusing is stronger. }
 \label{fig:sketch}
\end{figure}

The left panel of Fig.~\ref{fig:deltaTheta_vs_phi_GP_numerical} shows that the strength of the focusing strongly depends on the azimuthal angle of the electrons, being a factor of two smaller for electrons emitted at $\phi = \pi$ than for electrons emerging at $\phi = 0$. This dependence is a consequence of the crossing angle, as depicted in Fig.~\ref{fig:sketch}: a lepton emitted towards the inside of the ring, corresponding to $\phi = 0$,
travels for quite some time in the vicinity of the opposite charge bunch, and feels a stronger focusing force than a lepton emitted on the other side towards $\phi = \pi$, which is further away from the opposite charge bunch. The two curves shown in Fig.~\ref{fig:deltaTheta_vs_phi_GP_numerical} (left) correspond to different predictions: the numerical calculation described in Section~\ref{sec:NumericalCalc}, and the {\tt Guinea-Pig} simulation with the nominal settings mentioned in Section~\ref{sec:GuineaPig}
The result from the numerical integration agrees well with the {\tt Guinea-Pig} simulation, 
in particular for electrons that are emitted in, or close to, the $(x,z)$ plane of the collision. 
The dependence shown in Fig.~\ref{fig:deltaTheta_vs_phi_GP_numerical} (left) is used in Section~\ref{sec:CorrectionInsitu} to build an experimental observable that is strongly correlated with the luminosity bias.

The right panel of Fig.~\ref{fig:deltaTheta_vs_phi_GP_numerical} shows the deflections in $x$ and in $y$ separately, as a function of the azimuthal angle of the electrons, again for $45.6$\,GeV electrons emitted at $\theta^* = 64$\,mrad. The latter are defined as $\Delta x' = \left( p_x / E \right)_{0} - \left( p_{x} / E \right)_{\rm{final}}$, and similarly for $\Delta y'$. Particles emitted at $| \phi |< \pi/2$  ($ | \phi | > \pi/2$) 
have a positive (negative) momentum along the $x$ direction, such that a  positive (negative) value for $\Delta x'$ corresponds indeed to a focusing deflection. 
Similarly, the sign of $\Delta y'$ as seen in the figure corresponds to a focusing deflection along $y$.
While, for flat beams with $\sigma_{y} \ll  \sigma_{x}$ that collide head-on, the deflection would be primarily along the $y$ direction, the figure shows that, in the presence of a crossing angle, the deflection in the $x$ direction also plays an important role.

\subsubsection{Dependence on the settings of the simulation}

\begin{figure}[htbp]
\begin{center}
\begin{tabular}{cc}
\includegraphics[width=0.5\columnwidth]{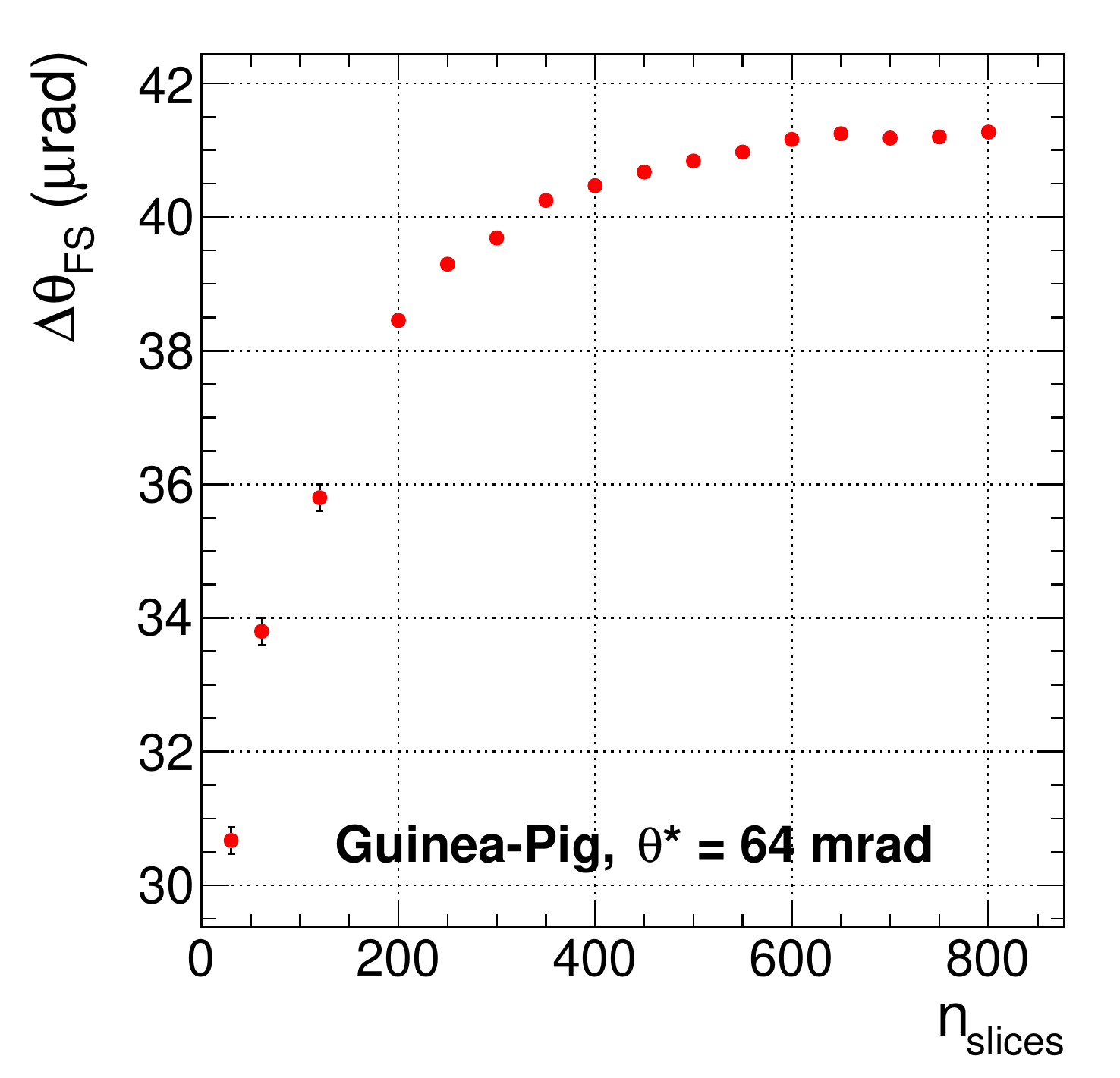} &
\includegraphics[width=0.5\columnwidth]{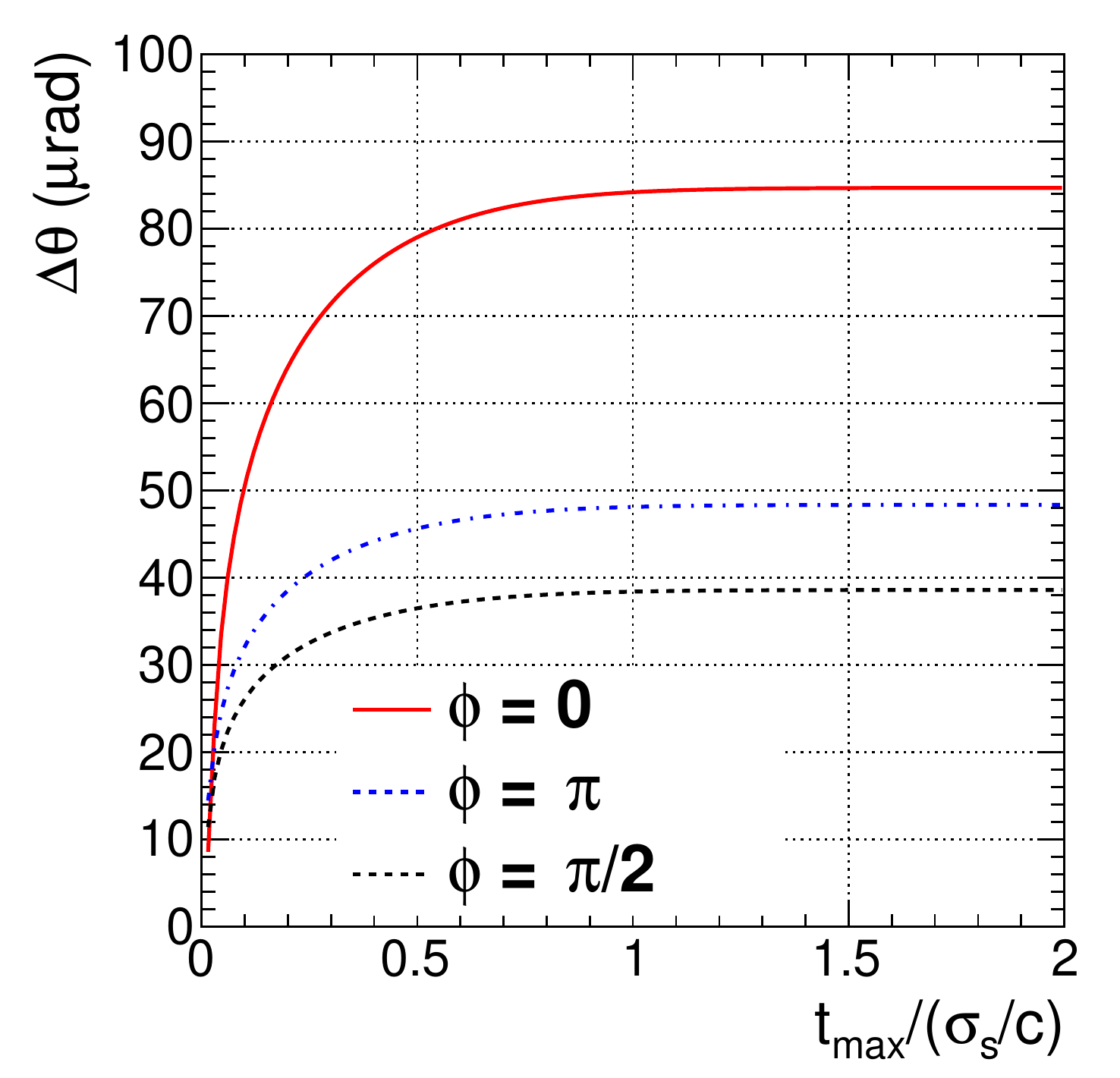} 
\end{tabular}
\caption{\small Left: Average deflection angle as predicted by {\tt Guinea-Pig} as a function of the number of longitudinal slices of the bunch. Right: Deflection angle of an electron with $E = 45.6$\,GeV emerging at $\theta^* = 64$\,mrad from a Bhabha interaction taking place at the nominal interaction point and at $t=0$, as a function of time, for three example values of the azimuthal angle of the electron.
}
\label{fig:deltaTheta_buildup}
\end{center}
\end{figure}

The angular focusing $\Delta \theta_{\rm{FS}}$ of electrons emerging from a Bhabha interaction was found to be more sensitive than the $p_x$ kick to the settings of the {\tt Guinea-Pig} simulation. The left panel of Fig.~\ref{fig:deltaTheta_buildup} shows the {\tt Guinea-Pig} prediction for the angular focusing of $45.6$\,GeV electrons emitted at $\theta^* = 64$\,mrad, when the number of longitudinal slices is increased from $30$ to $800$, the other settings being identical to the default settings given in Section~\ref{sec:GuineaPig}. The convergence is seen to be reached with about $700$ slices, which corresponds to a slice length of about $20\%$ of the 
standard deviation $\Sigma_z$ of the $z_{\rm{vtx}}$ distribution (Eq.~\ref{eq:SigmaIR}).

The right panel of Fig.~\ref{fig:deltaTheta_buildup} shows the deflection angle $\Delta \theta$ of a $45.6$\,GeV electron emerging at $\theta^* = 64$\,mrad from a Bhabha interaction taking place at $t=0$ and at a spatial vertex corresponding to the nominal interaction point. The deflection is obtained by integrating the Lorentz force felt by the electron between $t=0$ and a time $t_{\rm{max}}$ shown on the $x$ axis, expressed in units of $\sigma_{s} / c$. It can be seen that the deflection angle quickly reaches its plateau value $\Delta \theta_{\rm{FS}}$, at a time of about $0.7 \sigma_{s}/c$, irrespective of the azimuthal angle of the electron. This means that the deflection of Bhabha electrons from the field of the opposite charge bunch remains a rather localised effect. In particular, when the electron is emitted in the $(x,z)$ plane, hence at $49$\,mrad ($79$\,mrad) with respect to the $z$ axis for $\phi = 0$ ($\phi = \pi$), it has travelled a distance of $0.4$\,mm ($0.7$\,mm) along the $x$ direction at $t = 0.7 \sigma_{s} / c$. Consequently, when {\tt Guinea-Pig} is used to determine $\Delta \theta_{\rm{FS}}$, the knowledge of the fields within the ``first'' grid, set to extend up to $150 \sigma_x \simeq 1$\,mm along the $x$ direction, is sufficient for an accurate determination of the deflection angle. When the electron is emitted in the $(y,z)$ plane, at $\phi = \pi/2$, it reaches $y = 0.5$\,mm at $t = 0.7 \sigma_{s} / c$. This distance is very large compared to the dimension along $y$ of the first grid, set to $60 \sigma_y \simeq 2\,\mu$m. However, as can be seen in Fig.~\ref{fig:deltaTheta_vs_phi_GP_numerical} (left), the approximation made when {\tt Guinea-Pig} is run with one single grid, whereby the fields outside the grid are taken to be those expected from a linear charge distribution, remains reasonable also in that extreme case, as the {\tt Guinea-Pig} prediction agrees within less than $10\%$ with the numerical calculation. 
This agreement justifies the choice of the default {\tt Guinea-Pig} settings used for this study.  \\

\subsubsection{Determination of the luminosity bias and its dependence with respect to the beam parameters}

\begin{figure}[bhtp]
\begin{center}
\begin{tabular}{cc}
\includegraphics[width=0.5\columnwidth]{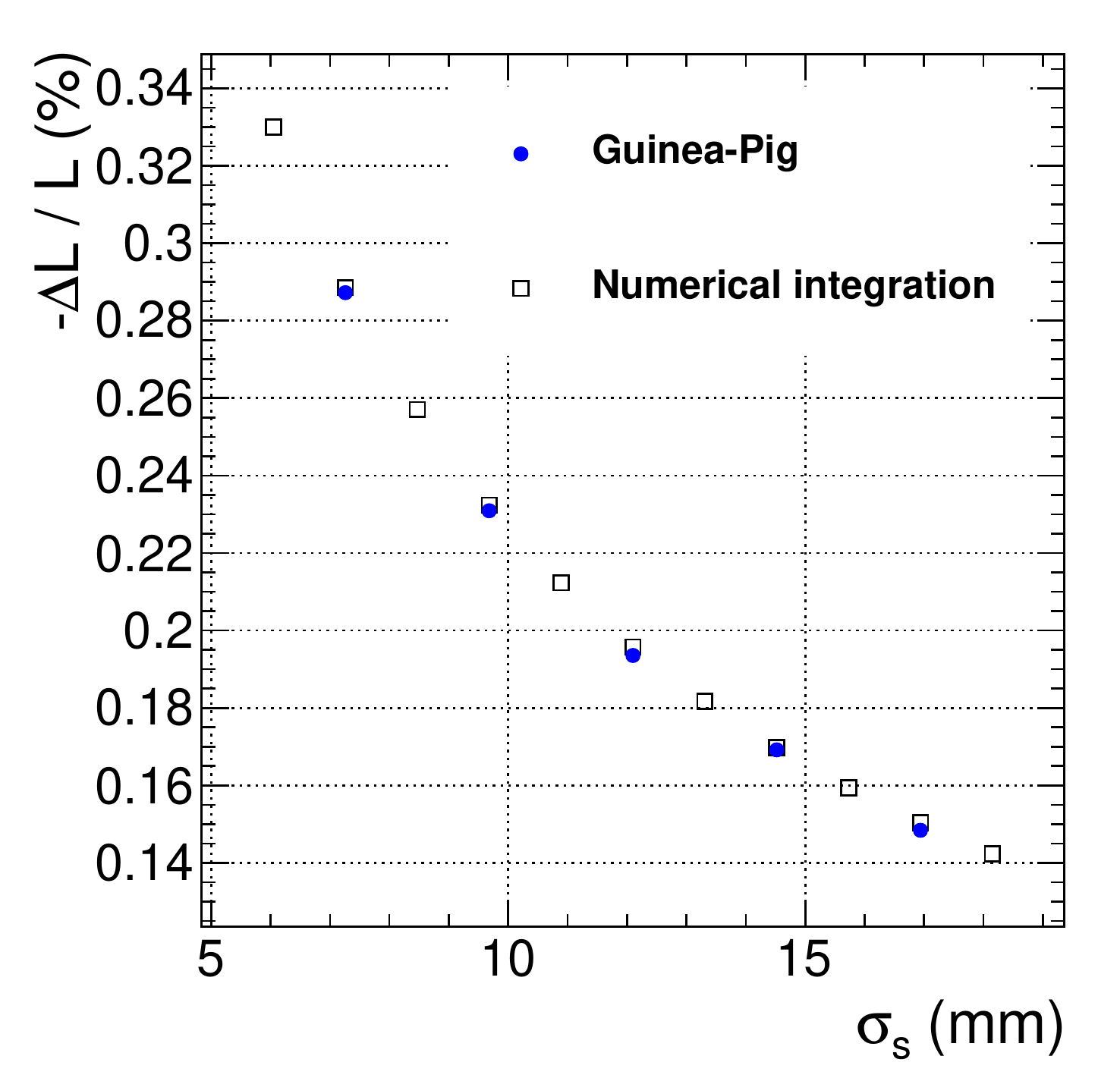} &
\includegraphics[width=0.5\columnwidth]{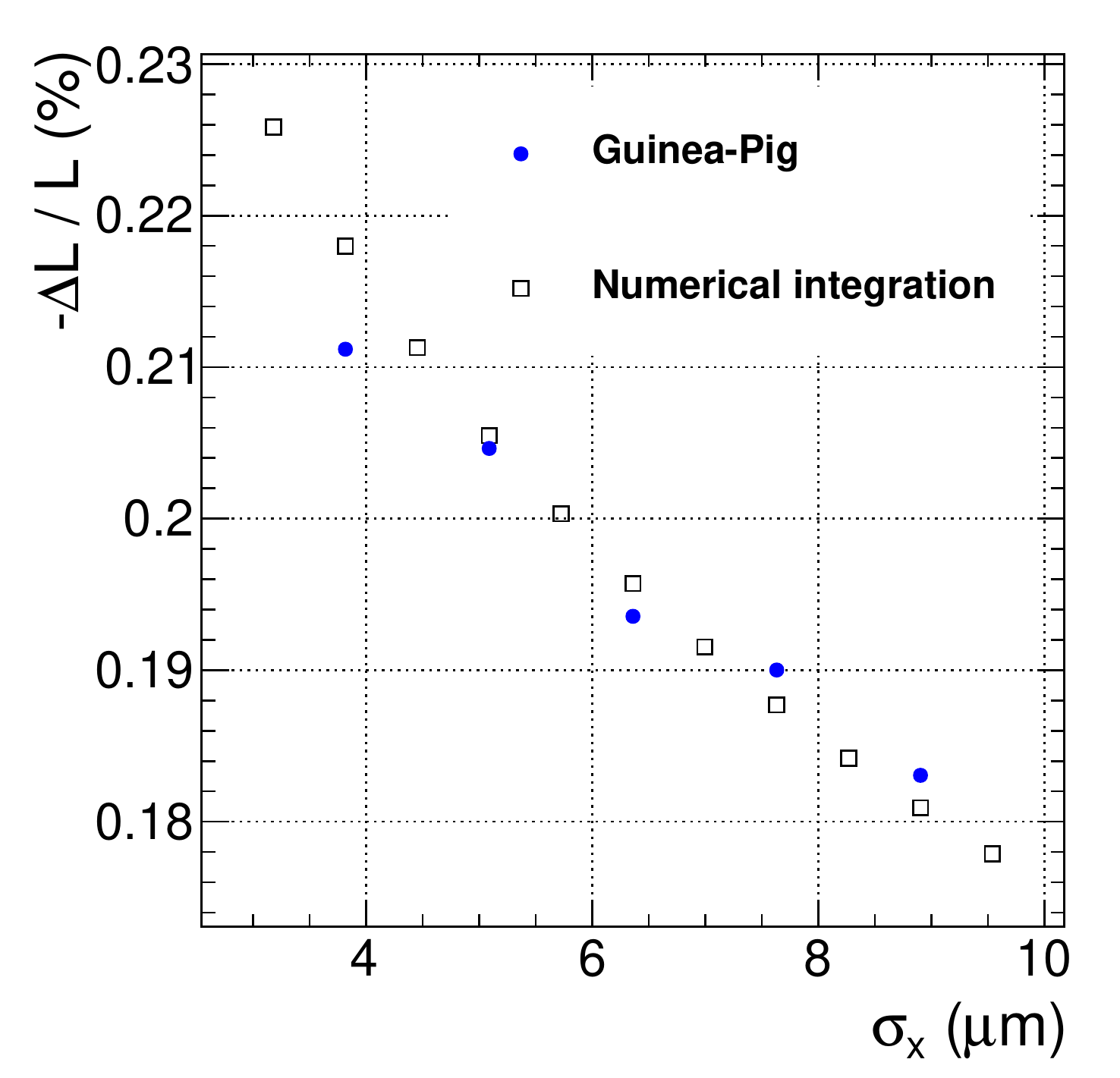}
\end{tabular}
\caption{\small Variation of the luminosity bias as a function of (left) the longitudinal bunch length and (right) the transverse bunch size along the $x$ direction, as predicted by the {\tt Guinea-Pig} simulation (closed dots) and by a numerical integration of the average force felt by the electrons (open squares). 
}
\label{fig:LumiBias_vs_params}
\end{center}
\end{figure}

In all what follows, the luminosity bias (as well as the observable described in Section~\ref{sec:CorrectionInsitu}) is determined for ``leading-order Bhabha'' events, i.e.\ events in which an electron and a positron, of $45.6$\,GeV each, are emitted back-to-back in the centre-of-mass frame of the collision, with an angle $\theta^*$ distributed according to $ 1 / \theta^{*3}$. Under this approximation, the observable and the luminosity bias can be calculated numerically (e.g.\ from Eq.~\ref{eq:lumiBias} for the latter), once the $p_x$ kick and the final state angular focusing $\Delta \theta_{\rm{FS}}(\theta, \phi)$ are known, either from the {\tt Guinea-Pig} simulation or from the numerical calculation described in Section~\ref{sec:NumericalCalc}. 

A sample of $7$ millions of ``genuine'' Bhabha events, generated with the {\tt BHWIDE} Monte-Carlo program, are also used in Sections~\ref{sec:correlation_initial_final} and~\ref{sec:CorrectionInsitu} to determine the luminosity bias. Photon radiation (included in {\tt BHWIDE}) leads to softer electrons in the final state, which feel a stronger focusing  (Fig.~\ref{fig:deltaTheta_BHWIDE} right). For final state radiations, however, the photon is usually emitted at a small angle with respect to the final state electron. The clustering algorithm that will be used to reconstruct the electrons in the LumiCal is likely to merge the electron and the (non deflected) radiated photon into a single cluster, thereby compensating for the latter effect. Hence, a proper study of the effect of radiations requires the {\tt BHWIDE} events to be processed through a full simulation of the LumiCal and a cluster reconstruction algorithm to be run on the simulated energy deposits. Such a full simulation study, on a large statistics sample, is beyond the scope of this paper. Nonetheless, the luminosity bias (and the observable described in Section~\ref{sec:CorrectionInsitu}) obtained from {\tt BHWIDE} events are also shown below, as obtained using the final state charged leptons, applying a loose lower energy cut of $5$\,GeV on the latter, and ignoring the effects of the LumiCal clustering. 
The true values of the bias and of the observable
are expected to lie between this latter determination and the prediction corresponding to leading-order Bhabha events. \\

The dependence of the luminosity bias on the parameters that characterise the bunches has also been investigated.
As any other quantity related to the beam-beam effects considered here, the bias trivially scales linearly with the intensity of the  bunches (everything else being equal). The only other parameters for which a significant dependence has been seen are the longitudinal bunch length, and the transverse size of the bunches in the $x$ direction. Figure~\ref{fig:LumiBias_vs_params} shows how the luminosity  bias (as determined for leading-order Bhabha events)  depends on these two parameters. The {\tt Guinea-Pig} simulation and the numerical integration code predict very similar values, the largest difference between both predictions, observed when the transverse size $\sigma_x$ is lowered by $40 \%$ compared to its nominal value\footnote{The length $\Sigma_z$ (Equation~\ref{eq:SigmaIR}) being then reduced in the same proportion, the length of each of the $750$ slices used in the {\tt Guinea-Pig} simulation becomes larger than $30\%$ of $\Sigma_z$. Consequently, the beam-induced effects predicted by {\tt Guinea-Pig} are likely to be slightly underestimated in that case (Fig.~\ref{fig:deltaTheta_buildup} left).}, being smaller than $10^{-4}$. Varying the longitudinal bunch length by  $5\%$ around its nominal value modifies the luminosity bias by about $5\%$. A similar variation of the luminosity bias is observed when the transverse size $\sigma_x$ is varied by about $25 \%$ around its nominal value.



\subsection{Correlation between the effects on the initial and on the final state particles} 
\label{sec:correlation_initial_final}

\begin{figure}[!htb]
  \begin{center}
\begin{tabular}{cc}
\includegraphics[width=0.49\columnwidth]{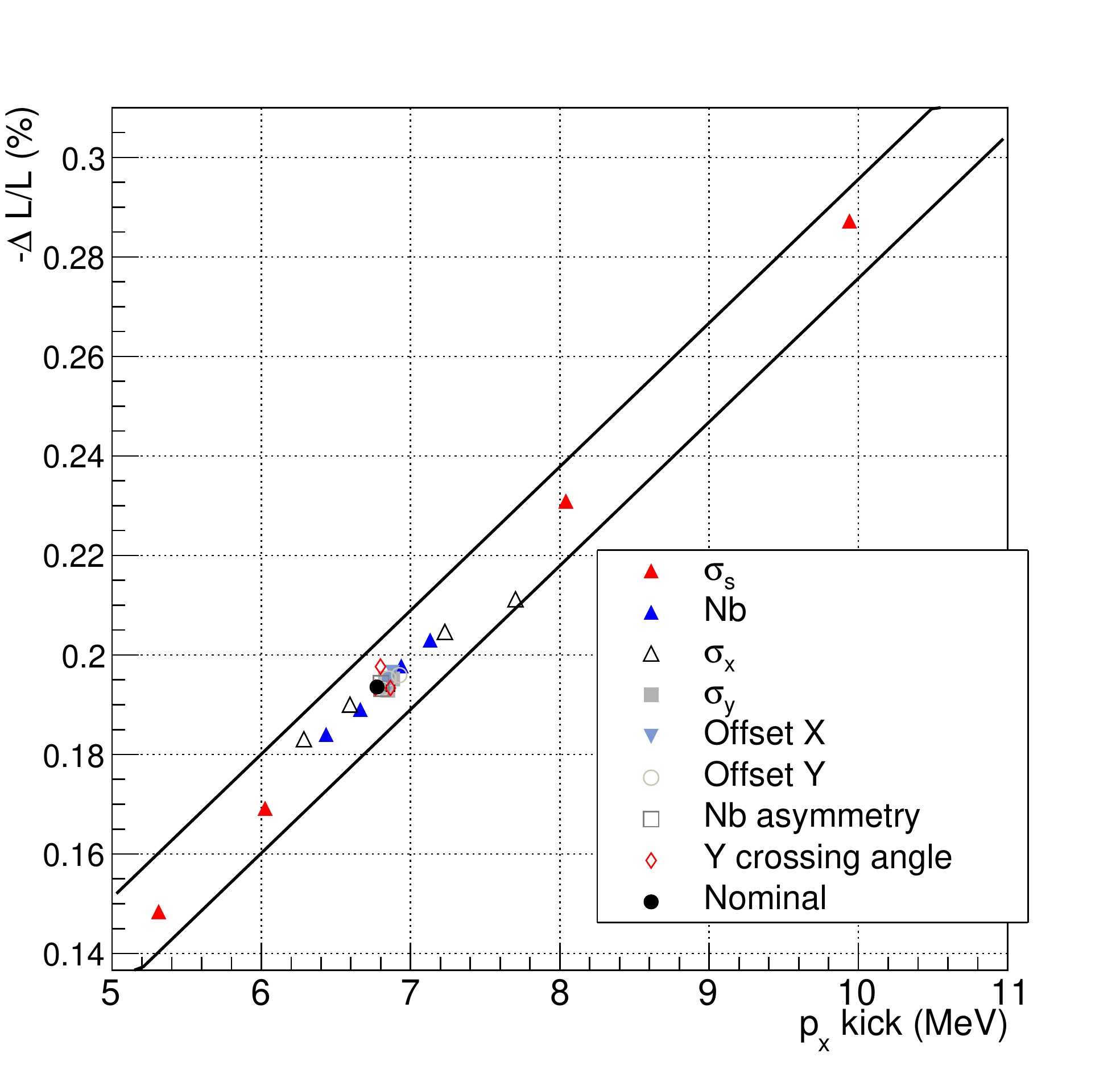} &
\includegraphics[width=0.49\columnwidth]{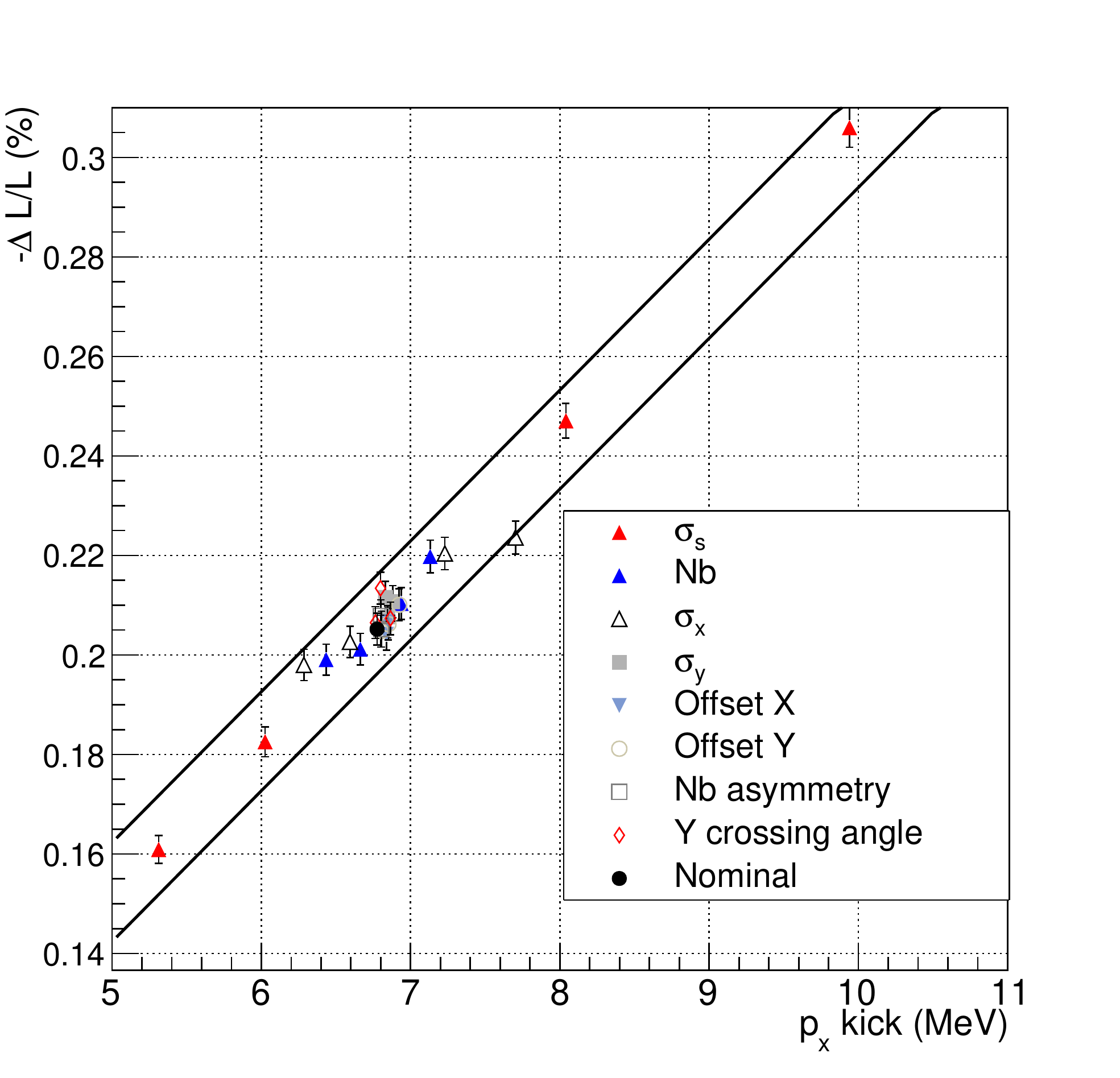}
\end{tabular}
  \caption{Luminosity bias as a function of the (absolute value of the) event kick (i.e.\ twice the $k_x$ of Section~\ref{sec:bbInitialState}), for (left)  leading-order Bhabha events and (right) from {\tt BHWIDE} events, determined for several variations of the beam parameters. The two  lines indicate a $\pm 10^{-4}$ variation around a linear fit to all points.}
  \label{fig:lumibias_versus_kick}  
  \end{center}
\end{figure}

A strong correlation is expected between the beam-beam effects in the initial state of an $\rm e^+ e^-$ interaction, and the beam-induced deflection of the charged leptons in the final state of a Bhabha event, since the source of both effects is identical. To check that this expectation is borne out by the numerical calculations, several scenarios of beam parameters have been considered, whereby one parameter is varied around its nominal value given in Table~\ref{tab:parameters} while the others remain fixed:
\begin{itemize}
\item the bunch intensity has been varied by $\pm 2 \%$ and $\pm 5 \%$; 
\item the longitudinal bunch length $\sigma_s$, as well as the bunch transverse sizes $\sigma_x$ and $\sigma_y$, by $\pm 20 \%$ and $\pm 40 \%$; 
\item an horizontal (vertical) relative beam offset has been set, equal to $20\%$ or $40\%$ of $\sigma_x$ ($\sigma_y$); 
\item a crossing angle $\alpha_y$ in the $(y, z)$ plane has been set with $\alpha_y / 2 = 10\,\mu$rad, $50\,\mu$rad and $100\,\mu$rad;
\item an asymmetry of $2\%$ and $5\%$ between the number of particles in the $\rm e^-$ and the $\rm e^+$ bunches has been set.
\end{itemize}
The latter variation accounts for the intrinsic asymmetry induced by the top-up injection scheme. For the other parameters, the range considered for these variations was deliberately chosen to be very large compared to the expected accuracy with which these parameters can be monitored~\cite{Abada2019}. For each scenario, the $p_x$ kick is determined, as well as the predicted bias on the luminosity measurement, and they are plotted against each other in Fig.~\ref{fig:lumibias_versus_kick}. The left panel of Fig.~\ref{fig:lumibias_versus_kick} shows this bias as determined for leading-order Bhabha events, while in the right panel, the bias obtained from a sample of $7$ millions of {\tt BHWIDE} events is shown. 
A comparison of both plots shows that, as expected, the luminosity bias increases slightly  when photon radiations are included, since softer electrons  experience a stronger deflection.
The luminosity bias is seen to be a linear function of the kick indeed, this function being independent of which parameter has been varied. All points remain within $\pm 10^{-4}$ of the prediction of a linear fit to all scenarios, even for the very large variations considered here. Consequently, once one knows the value of the $p_x$ kick, one knows, to the required precision, the factor by which the luminosity should be corrected to account for the beam-induced effects. 

This correlation between the beam-beam effects on the particles in the initial state of an $\rm e^+ e^-$ interaction and the luminosity bias is exploited in Sections~\ref{sec:CorrectionDimuons} and~\ref{sec:CorrectionInsitu} to determine the luminosity correction factor.

\section{Correction using the central detector}

\label{sec:CorrectionDimuons}

As seen in Section~\ref{sec:bbInitialState}, the $p_x$ kick causes an increase, $\delta E$, of the energy of the particles at the time when they interact, and an increase of the crossing angle from $\alpha_0$ to $\alpha = \alpha_0 + \delta\alpha$, given by 
\begin{equation}
 \delta \alpha  =  2 k_x  / E_{\rm beam}  \simeq 0.5 \% \times \alpha_0 ,
  \label{eqn:deltaalpha}
\end{equation}
where $\alpha_0$ is the nominal crossing angle. 
As shown in Ref.~\cite{blondel2019polarization}, a precise measurement of $\delta \alpha$ is a crucial ingredient for a precise determination of the centre-of-mass energy of the collision, the motivation being summarised in what follows. Since the kick has no component along the $z$-axis, it has no effect on the centre-of-mass energy of the collision, given by $$ \sqrt s = 2 \sqrt{ p_{z,+} p_{z,-}} = 2 \sqrt{E_+ E_-} \cos\alpha/2 ,$$ 
where $p_{z,-}$ ($p_{z,+}$) denote the absolute value of the momentum of the incoming electrons (positrons) along the $z$ axis, and $E_\pm = E_\pm^0+\delta E_\pm$ their energy at the time when they interact.
The method of resonant depolarisation of non-colliding bunches, which are not affected by beam-beam effects,  provides a very precise measurement of the nominal beam energies, $E_{\pm}^0$.
From this measurement, the centre-of-mass energy of the collision can be derived
as 
$$ \sqrt{s} = 2 \sqrt{E_+^0 E_-^0} \cos \alpha_0/2, $$ 
provided that the nominal crossing angle $\alpha_0$ is known. Since the effective crossing angle $\alpha = \alpha_0 + \delta \alpha$ can be measured precisely by exploiting the over-constrained kinematics of $\rm e^+ \rm e^- \rightarrow \rm{\mu}^+ \rm{\mu}^-$ events~\cite{blondel2019polarization}, the determination of the nominal crossing angle boils down to measuring the crossing angle increase $\delta \alpha$ that is induced by the beam-beam effects. A method has been proposed in Ref.~\cite{blondel2019polarization} to perform this measurement. It is shown therein that $\delta \alpha$ can be determined with a relative  accuracy of about $2 \%$ by 
measuring $\alpha$ from dimuon events during the various steps of the filling period of the machine, and extrapolating these measurements to the limit where the beam intensities (hence the beam-induced effects) vanish.
From Equation~\ref{eqn:deltaalpha}, this accuracy on $\delta \alpha$ directly translates into a relative uncertainty of $2 \%$ on the $p_x$ kick, and, from Fig.~\ref{fig:lumibias_versus_kick}, into the same uncertainty of $2 \%$ on the luminosity correction factor, well within the target precision. This determination fully relies on dimuon events measured in the central detector. 


\section{In-situ correction using the luminometer}

\label{sec:CorrectionInsitu}

A complementary method has also been developed, whereby, in contrast to the method described in Section~\ref{sec:CorrectionDimuons}, the luminosity correction factor can be determined in-situ, using only measurements made in the luminometer system. It relies on the definition of an experimental observable which
is largely driven by the $p_x$ kick and, as the latter, is strongly correlated with the luminosity bias.

\subsection{Acollinearity variable}

\begin{figure}[htbp]
\begin{center}
\begin{tabular}{cc}
\includegraphics[width=0.48\columnwidth]{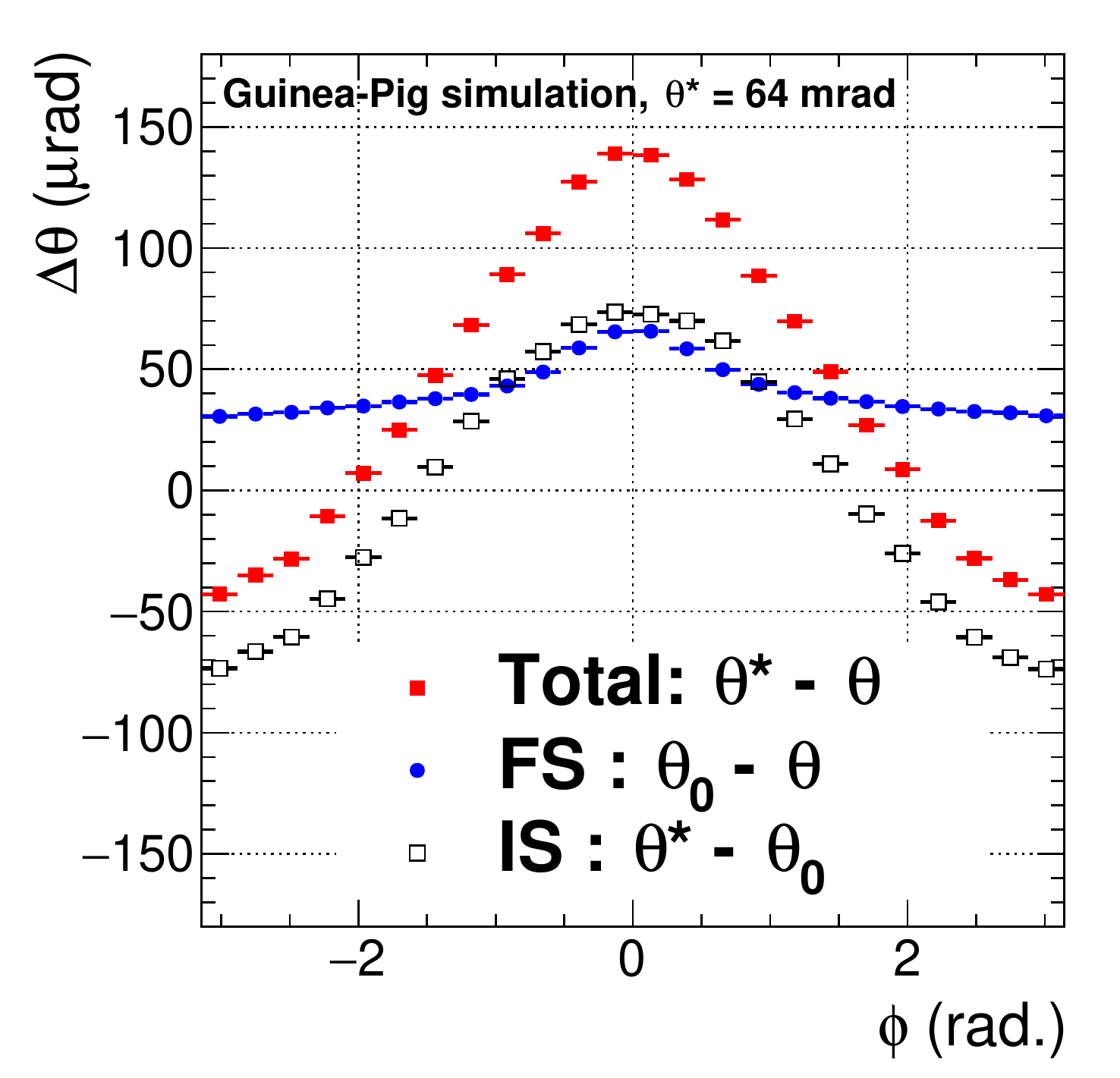} &
\includegraphics[width=0.48\columnwidth]{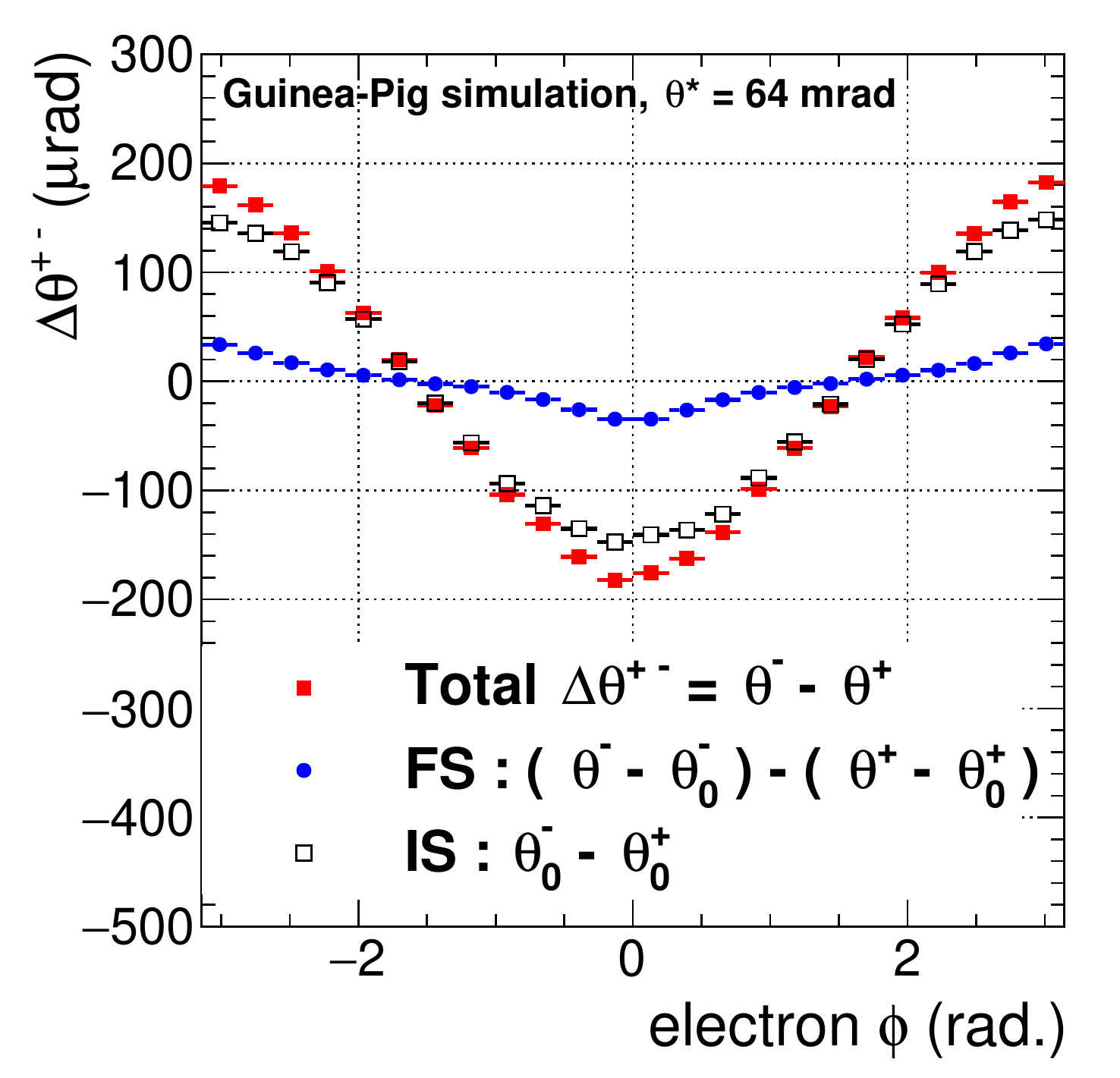}
\end{tabular}
\caption{\small Left: Angular focusing of a $45.6$\,GeV $\rm e^{\pm}$ emitted at an angle of $64$\,mrad as a function of its azimuthal angle. The combined effect of the $p_x$ kick acquired by the initial state (IS, black open squares) and of the focusing of the final state (FS) lepton as it emerges from the interaction (blue closed dots) is shown as the red closed squares. Right: Acollinearity of the final state of a Bhabha event, defined as the difference $\theta^- - \theta^+$ where $\theta^-$ ($\theta^+$) denotes the polar angle of the $\rm e^-$ ($\rm e^+$) with respect to the electron (positron) beam direction, as a function of the azimuthal angle of the electron. The acollinearities induced by the initial state and by the final state effects are also shown separately. 
}
\label{fig:observables_vs_phi}
\end{center}
\end{figure}

The azimuthal dependence of the beam-beam effects described earlier is exploited to define the aforementioned observable. The  effects of the $p_x$ kick (Fig.~\ref{fig:deltaTheta_and_phi_kick-GP}, middle, and curve labelled ``IS'', for initial state, in Fig.~\ref{fig:observables_vs_phi}, left) and of the focusing of the final state $\rm e^{\pm}$ by the opposite charge bunch (Fig.~\ref{fig:deltaTheta_vs_phi_GP_numerical}, and curve labelled ``FS'', for final state, in Fig.~\ref{fig:observables_vs_phi}, left) add up and the total effect is shown by the closed squares in Fig.~\ref{fig:observables_vs_phi}, left: while $45.6$\,GeV $\rm e^{\pm}$ emitted at $\theta^* = 64$\,mrad and $\phi = 0$ are focused by about $150\,\mu$rad, they are deflected towards larger angles (``defocusing'') by about $50\,\mu$rad when emitted at $\phi = \pi$. 
The different deflections felt by two particles that are separated by $\pi$ in azimuth lead to an acollinearity of the final state of a Bhabha interaction: the difference $\Delta \theta^{+ \, -} = \theta^- - \theta^+$ between the polar angle of the electron, $\theta^-$, and that of the positron, $\theta^+$, both measured with respect to the direction of the respective beam, is non-vanishing and strongly depends on the azimuthal angle of (for example) the electron, as shown in the right panel of Fig.~\ref{fig:observables_vs_phi}. The observable used here is an explicit measure of the modulation of this acollinearity.
It is built
from the averages of $\Delta \theta^{+ -}$ in $ | \phi^- | < \pi / 2$ and in $ | \phi^- | > \pi /2$, $\phi^-$ denoting the azimuthal angle of the electron. These two quantities are expected to be opposite, and, by definition, are measured with independent events. We define the variable {\tt Acol} as:
\begin{equation}
 {\tt Acol} =  \left < \Delta \theta^{+ -} \right>_{  | \phi^- | > \pi / 2}  - \left < \Delta \theta^{+ -} \right>_{ | \phi^- | < \pi / 2} \qquad . 
 \label{eq:acol_def}
\end{equation}

For the nominal configuration at the $\rm Z$ pole, the {\tt Guinea-Pig} simulation predicts that {\tt Acol} is about $217\,\mu$rad for leading-order Bhabha events, $190\,\mu$rad being induced by the $p_x$ kick and $27\,\mu$rad being due to the final state deflections.
Within each $\phi^-$ hemisphere, the RMS of the distribution of the acollinearity $\Delta \theta^{+ -} $ amounts to about $100\,\mu$rad for leading-order events. The resolution of the polar angle measurement in the LumiCal  smears  the $\Delta \theta^{+ -} $ distribution further. To estimate the latter, a {\tt GEANT4}~\cite{Agostinelli:2002hh} simulation of the response of the LumiCal described in Ref.~\cite{Abada2019} was performed 
and the clustering algorithm implemented in the software of the FCAL collaboration~\cite{Sadeh:2010ey} was used. A resolution of about $140\,\mu$rad on the polar angle of an electron measured in the LumiCal was obtained.
With a total RMS of the $\Delta \theta^{+ -} $ distribution of $250\,\mu$rad, $n$ Bhabha events measured in each $\phi^-$ hemisphere provide a measurement of $ \left < \Delta \theta^{+ -} \right>$ with an uncertainty of $250\,\mu{\rm rad} / \sqrt{n}$, the error on {\tt Acol} being a factor of $\sqrt{2}$ larger. Consequently, only a few hundreds of events are sufficient to ensure a measurement of {\tt Acol} with a relative uncertainty of $5\%$.

\begin{figure}[htbp]
\begin{center}
\begin{tabular}{cc}
\includegraphics[width=0.48\columnwidth]{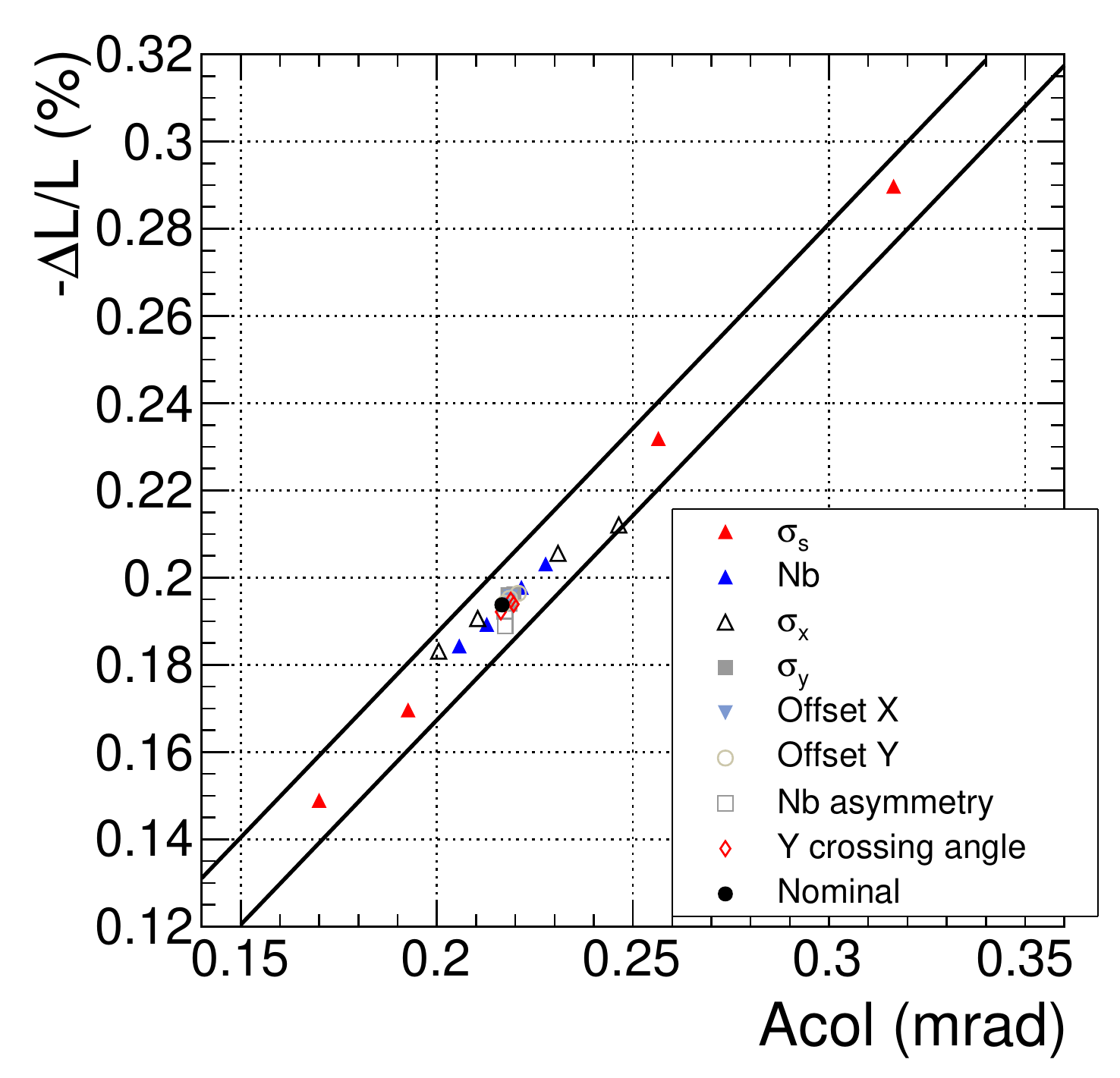} &
\includegraphics[width=0.48\columnwidth]{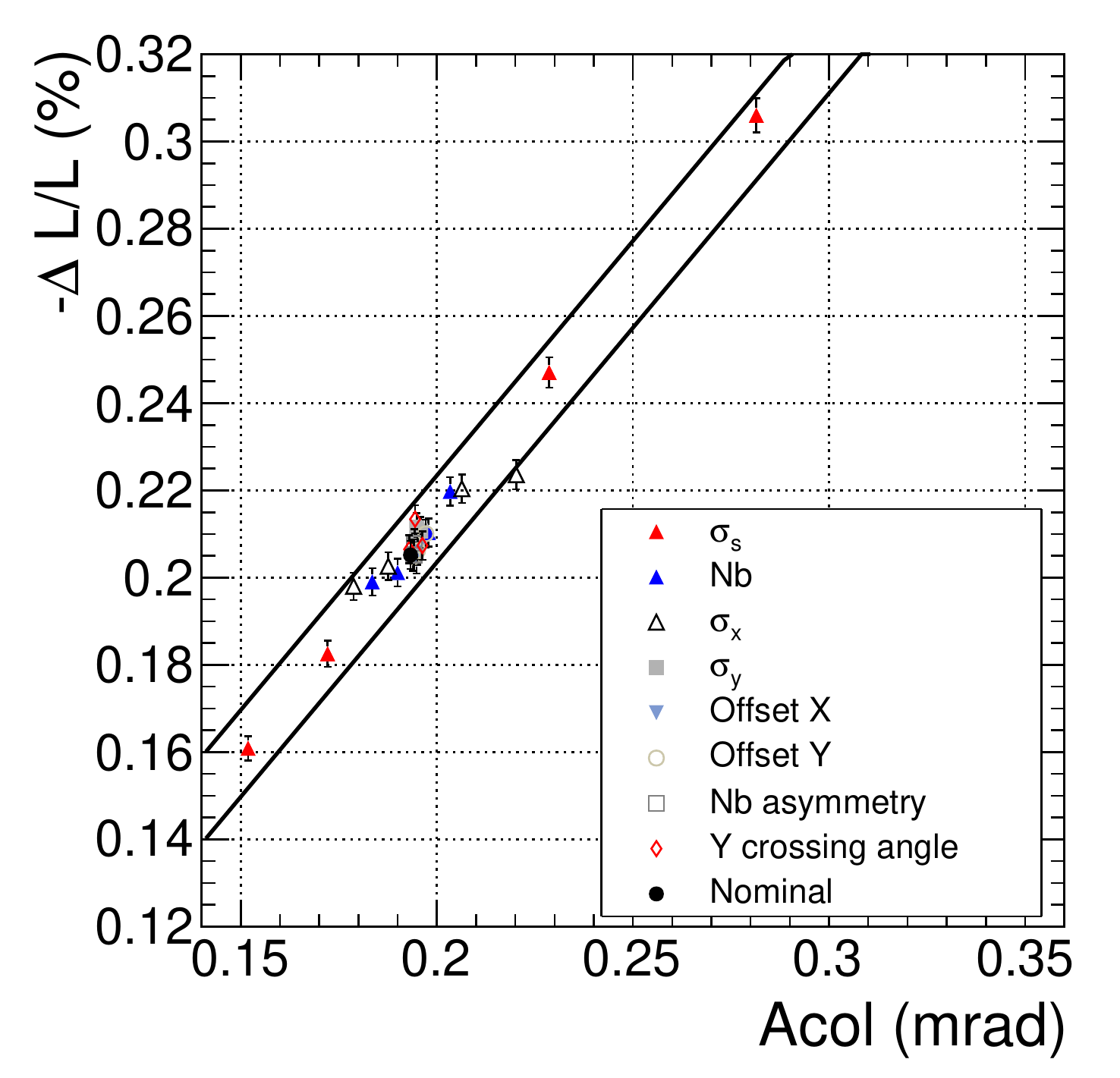}
\end{tabular}
\caption{\small Luminosity bias as a function of the {\tt Acol} variable for (left) leading-order Bhabha events and (right) from BHWIDE events. The two solid lines show a $\pm10^{-4}$ uncertainty band around a linear fit to all data points.
}
\label{fig:lumibias_vs_acolinearity}
\end{center}
\end{figure}

As the size of {\tt Acol} reflects the size of the beam-induced effects, it is expected to be strongly correlated with the luminosity bias. Figure~\ref{fig:lumibias_vs_acolinearity} shows that this is indeed the case. As  in Section~\ref{sec:correlation_initial_final}, the luminosity bias and {\tt Acol} have been determined for many sets of beam parameters, both in the case of leading-order Bhabha events (Fig.~\ref{fig:lumibias_vs_acolinearity}, left) and in the case of  Bhabha events generated by {\tt BHWIDE} (Fig.~\ref{fig:lumibias_vs_acolinearity}, right). Again, for all scenarios considered, which span a very large range of variations, the predicted bias on the luminosity lies inside a $\pm 10^{-4}$ uncertainty band around a linear fit to all points.
Consequently, the knowledge of {\tt Acol} provides a determination of the correction factor to be applied to the luminosity, with the desired precision.

A comparison between the left and right panels of Fig.\ref{fig:lumibias_vs_acolinearity} shows that the {\tt Acol} variable is lower for {\tt BHWIDE} events than for leading-order Bhabha events.
This effect is understood to be due to the intrinsic acollinearity induced by photon radiation, that smears the $\Delta \theta^{+ -}$ distribution.
As mentioned above, this difference between 
Bhabha events from BHWIDE 
and leading-order Bhabha events is expected to decrease when taking into account the effects of the clustering, and the actual relation between the luminosity bias and the {\tt Acol}  variable can be determined from dedicated simulations.

\subsection{Measurement of {\tt Acol} }

The acollinearity variable has two beam-induced components, the first being due to the beam-beam effects in the initial state, the second to the electromagnetic deflection of the final state particles. The component induced by the $p_x$ kick largely dominates (Fig.~\ref{fig:observables_vs_phi}). 

As mentioned in Section~\ref{sec:bbInitialState}, should the luminometer system be misaligned with respect to the IP along the $x$ direction, the $\theta$ distribution of the Bhabha electrons detected in the LumiCal would follow a modulation with $\phi$, similar to that caused by the $p_x$ kick. Consequently, such a misalignment $\delta_x$ causes another component to {\tt Acol}, {\tt Acol}$_{\rm{misalign}}$, which is proportional to $\delta_x$ and adds to the beam-induced component {\tt Acol}$_{\rm{beam}}$. While the method of asymmetric acceptance would ensure that the bias on the luminosity acceptance remains below $10^{-4}$ even for a large misalignment of $0.5$\,mm~\cite{Abada2019}, 
$\delta_x$ would have to be smaller than $5\,\mu$m for {\tt Acol}$_{\rm{misalign}}$ to be less than $5 \%$ of the nominal {\tt Acol}$_{\rm{beam}}$. 
Should such a level of alignment not be achieved, one would need to disentangle {\tt Acol}$_{\rm{beam}}$ from {\tt Acol}$_{\rm{misalign}}$ in order to derive the luminosity correction from the ``mapping'' shown in Fig.~\ref{fig:lumibias_vs_acolinearity}.
This distinction can be done by exploiting the fact that {\tt Acol}$_{\rm{beam}}$ scales linearly with the number of particles in the bunches, since both the $p_x$ kick and the angular deflection of the final state Bhabha electrons are proportional to the Lorentz force created by the beam, hence to the bunch intensity $N$. In contrast, {\tt Acol}$_{\rm{misalign}}$ is independent of $N$. A linear fit to  measurements of {\tt Acol}, made in bunches that  differ in intensity, allows in principle the intercept ({\tt Acol}$_{\rm{misalign}}$) and the slope ({\tt Acol}$_{\rm{beam}}$) to be determined.

\begin{figure}[htbp]
\begin{center}
\includegraphics[width=0.9\columnwidth]{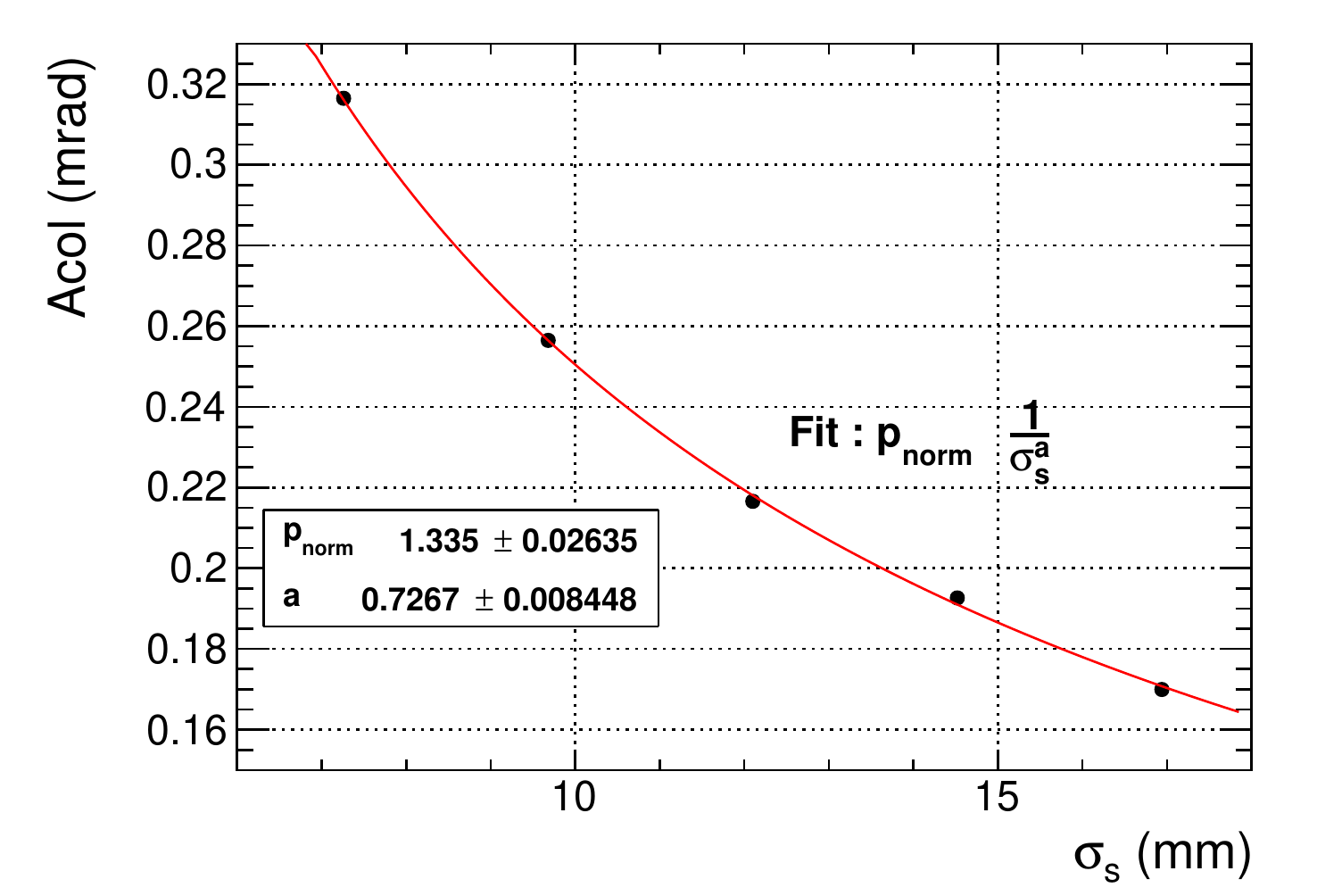}
\caption{\small The acollinearity variable as a function of the longitudinal length of the bunch, as determined from {\tt Guinea-Pig}, for leading-order Bhabha events.
The curve shows the result of a fit to a power law. 
}
\label{fig:asym_versus_sigmaz}
\end{center}
\end{figure}

However, colliding bunches that have a lower than nominal intensity suffer from less beamstrahlung  than the nominal bunches, and the bunch length,
which is largely driven by the length increase induced by beamstrahlung, is smaller than the nominal length $\sigma_s$ given in Table~\ref{tab:parameters}. Knowing how {\tt Acol} depends on $\sigma_s$, from the numerical calculations, allows this effect to also be accounted for. As can be seen from Fig.~\ref{fig:asym_versus_sigmaz}, this dependence can be approximated by a power law, ${\tt Acol} \sim 1 / \sigma^a_{s}$ with $ a \sim 0.72$.

\subsubsection{Using the ramp-up of the machine}

The filling period of the machine with the ``bootstrapping'' method naturally offers collisions with bunches that have a lower than nominal intensity. The idea of making measurements during this period and of extrapolating them to the situation where beam-beam effects would be absent has been proposed in Ref.~\cite{blondel2019polarization} and applied to the measurement of the crossing angle increase induced by the $p_x$ kick (Section~\ref{sec:CorrectionDimuons}). The same idea is exploited here. During this period, half of the nominal intensity is first injected in electron and positron bunches, which are then alternatively topped up by steps of $10 \%$, until the nominal intensity is reached for both. The bunches collide during this filling period, with the nominal optic parameters,
and their longitudinal lengths vary between $\sim 0.65 \sigma_s$ and $\sim 1.10 \sigma_s$~\cite{blondel2019polarization}.
The orbit may slightly differ from the nominal collision orbit, with a non-vanishing relative offset of the beams at the IP or a non-vanishing crossing angle in the vertical plane, but the {\tt Acol} variable is largely insensitive to such variations (Fig.~\ref{fig:lumibias_vs_acolinearity}).  Moreover, potential relative displacements of the beam spot between the ramp-up steps can be monitored precisely using tracks reconstructed in the tracker, such that a potential difference in {\tt Acol}$_{\rm{misalign}}$ between these steps can be corrected for.

\begin{figure}[tbp]
\begin{center}
\includegraphics[width=0.8\columnwidth]{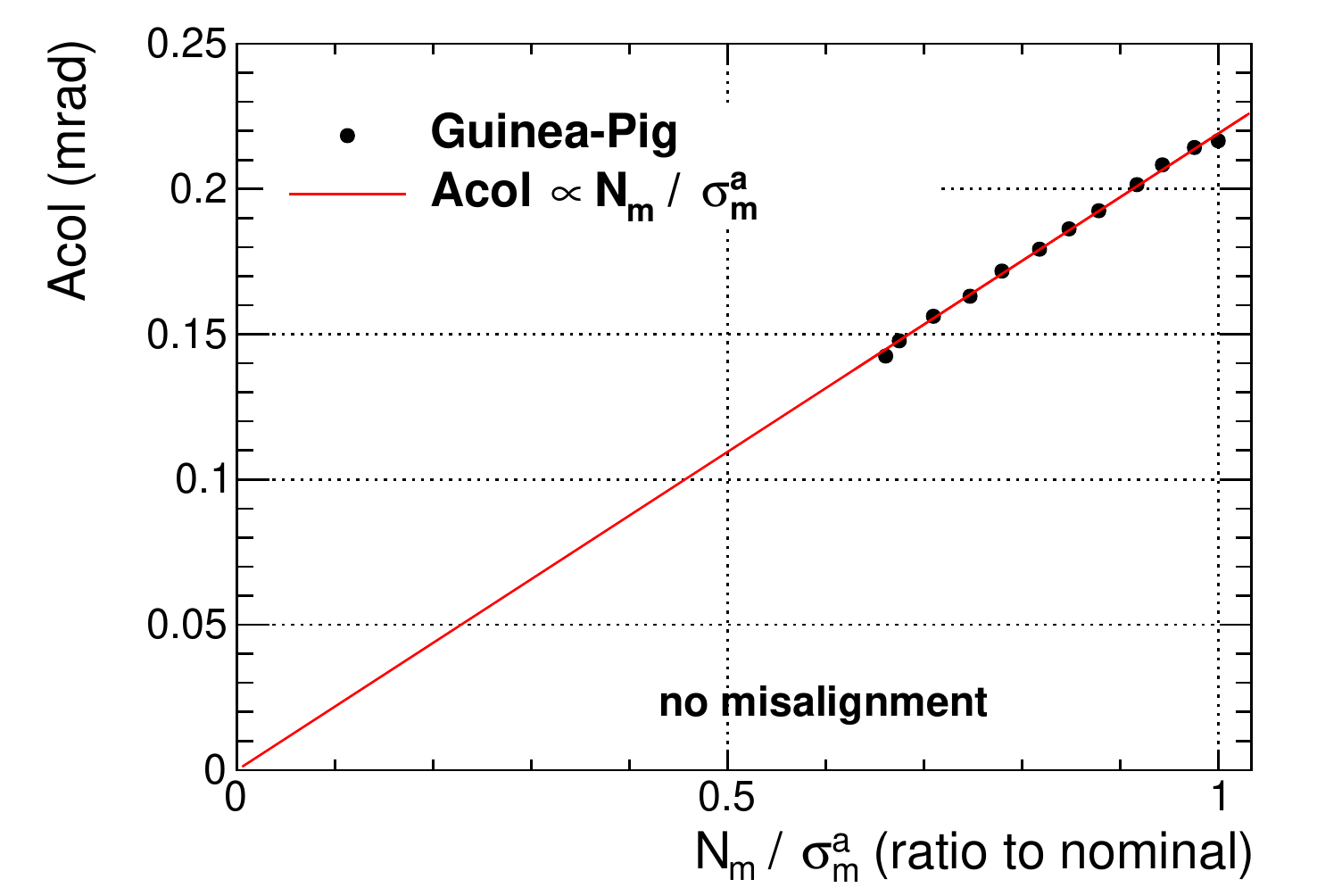} 
\caption{\small The acollinearity variable {\tt Acol}  predicted by {\tt Guinea-Pig} simulations at each step of the ramp-up, as a function of the average bunch intensity divided by the average bunch length to the power $0.72$, for leading-order Bhabha events. The result of a fit of an affine function is shown as the full line.
}
\label{fig:asym_rampup_GuineaPig}
\end{center}
\end{figure}

The observable {\tt Acol} is expected to scale approximately with $ N_{m} / \sigma_{m}^a$, where $ N_{m} \equiv \sqrt{ N_- N_+}$ with $N_-$ ($N_+$) denoting the intensity of the $\rm e^-$ ($\rm e^+$) bunches,  $\sigma_{m}$ is given by $\sigma_{m}^2 = ( \sigma^2_- + \sigma^2_+) / 2$ with $\sigma_-$ ($\sigma_+$) denoting their longitudinal length, and $a \simeq 0.72$ accounts for the $\sigma_s$ dependence of {\tt Acol}. 
This scaling has been checked explicitly by using the {\tt Guinea-Pig} program to determine the value of {\tt Acol}  that is expected during each filling step, for the nominal parameters given in Table~\ref{tab:parameters}, apart from the  intensities $N_-$ and $N_+$ and the 
bunch lengths $\sigma_-$ and $\sigma_+$, that were taken from Ref.~\cite{blondel2019polarization}. Figure~\ref{fig:asym_rampup_GuineaPig} shows that the results from {\tt Guinea-Pig} agree well with the expectation. The alignment of all points along a line that passes through the origin confirms the good description of the $\sigma_s$ dependence of {\tt Acol} by the chosen power-law (Fig.~\ref{fig:asym_versus_sigmaz}), in the $\sigma_s$ range of interest. The slope of this line is equal to the value of {\tt Acol} corresponding to the nominal intensities.

\begin{figure}[tbp]
\begin{center}
\includegraphics[width=0.8\columnwidth]{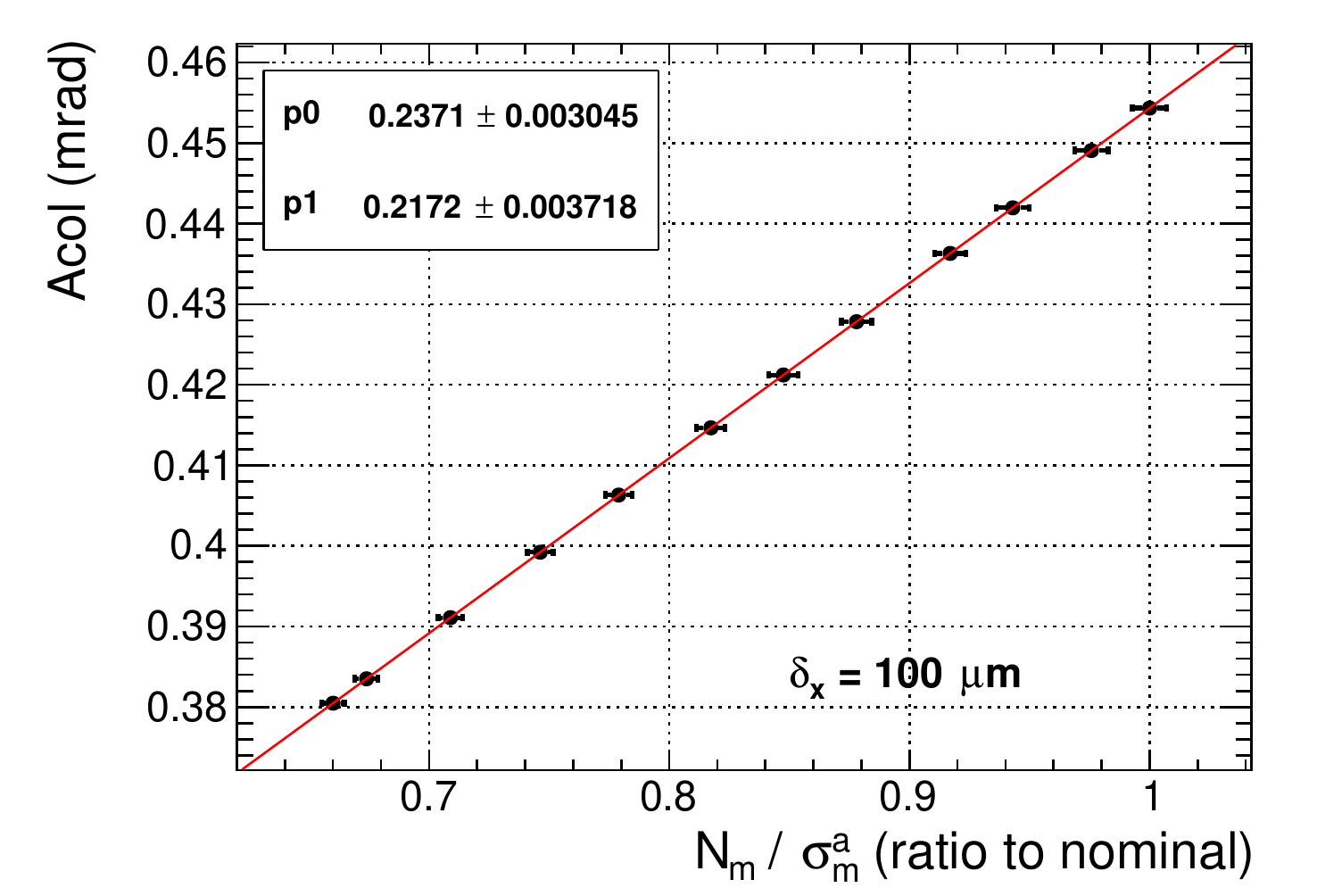} 
\caption{\small Acollinearity {\tt Acol} expected for leading-order Bhabha events at each step of the ramp-up, in presence of a misalignment of $100\,\mu$m of the luminometer system along the $x$ direction, as a function of the average bunch intensity divided by the average bunch length to the power of $a=0.72$. The errors on the points represent the statistical uncertainties obtained from $40$ seconds  of data at each point. The result of a linear fit is overlaid and the fitted slope ($p1$) and intercept ($p0$) are shown together with their uncertainties, the fitted values being equal to {\tt Acol}$_{\rm{beam}}$ and to the input {\tt Acol}$_{\rm{misalign}}$ by construction.
}
\label{fig:asym_rampup}
\end{center}
\end{figure}

To illustrate how the ramp-up of the machine can be used to disentangle the beam-induced and misalignment-induced components of {\tt Acol}, a misalignment of the luminometer system of $100\,\mu$m in the $x$ direction was assumed. 
This value would correspond to {\tt Acol}$_{\rm{misalign}} = 237\,\mu$rad, 
slightly larger than the beam-induced component of $217\,\mu$rad
predicted at the nominal intensities. 
For each filling step, the expected value of {\tt Acol} is taken to be $$ {\tt Acol} = {\tt Acol}_{\rm{misalign}} + {\tt Acol}_{\rm{beam}} \cdot \frac{  N_{m}  / \sigma_{m}^a  }  { \left( N / \sigma_s^a \right)_{nominal} } \qquad.$$ 
The  values for {\tt Acol} at each step
are shown in Fig.~\ref{fig:asym_rampup}, as a function of $N_{m} / \sigma_{m}^a$ normalised to its nominal value. The (very small) errors of {\tt Acol}  correspond to the statistics accumulated in $40$ seconds at each point.  The horizontal uncertainties on each point correspond to a relative uncertainty of $1\%$ on the bunch length, as sub-ps resolution should be obtained from bunch length monitoring measurements~\cite{Abada2019}. The uncertainty on the measurement of the bunch intensities is assumed to be negligible.  The result of a linear fit is also shown, from which it can be seen that the slope {\tt Acol}$_{\rm{beam}}$ could be determined with a statistical uncertainty of $1.7 \%$. This error would be reduced to less than $1 \%$ by using, instead of the intensities and bunch lengths provided by the beam monitoring, the number of dimuon events and the energy spread measured in-situ with a very good precision~\cite{blondel2019polarization}. 
The uncertainty due to the $\sigma_s$ dependence of {\tt Acol} is assessed by setting an uncertainty of $\pm 0.03$ for $a$ (which is conservative for the range of $\sigma_s$ considered here). It results in a systematic uncertainty of $2 \%$ on the fitted slope, well within the level of precision that is targeted for. Moreover, as long as the experimental resolution on the polar angle measurement dominates the spread of the $\Delta \theta^{+ -}$ distributions, the  statistical uncertainty of $1.7 \%$ on the luminosity correction factor, resulting from a fit to the {\tt Acol} measurements, is independent of the misalignment.

\subsubsection{Using pilot bunches}

\begin{figure}[htbp]
\begin{center}
\begin{tabular}{cc}
\includegraphics[width=0.53\columnwidth]{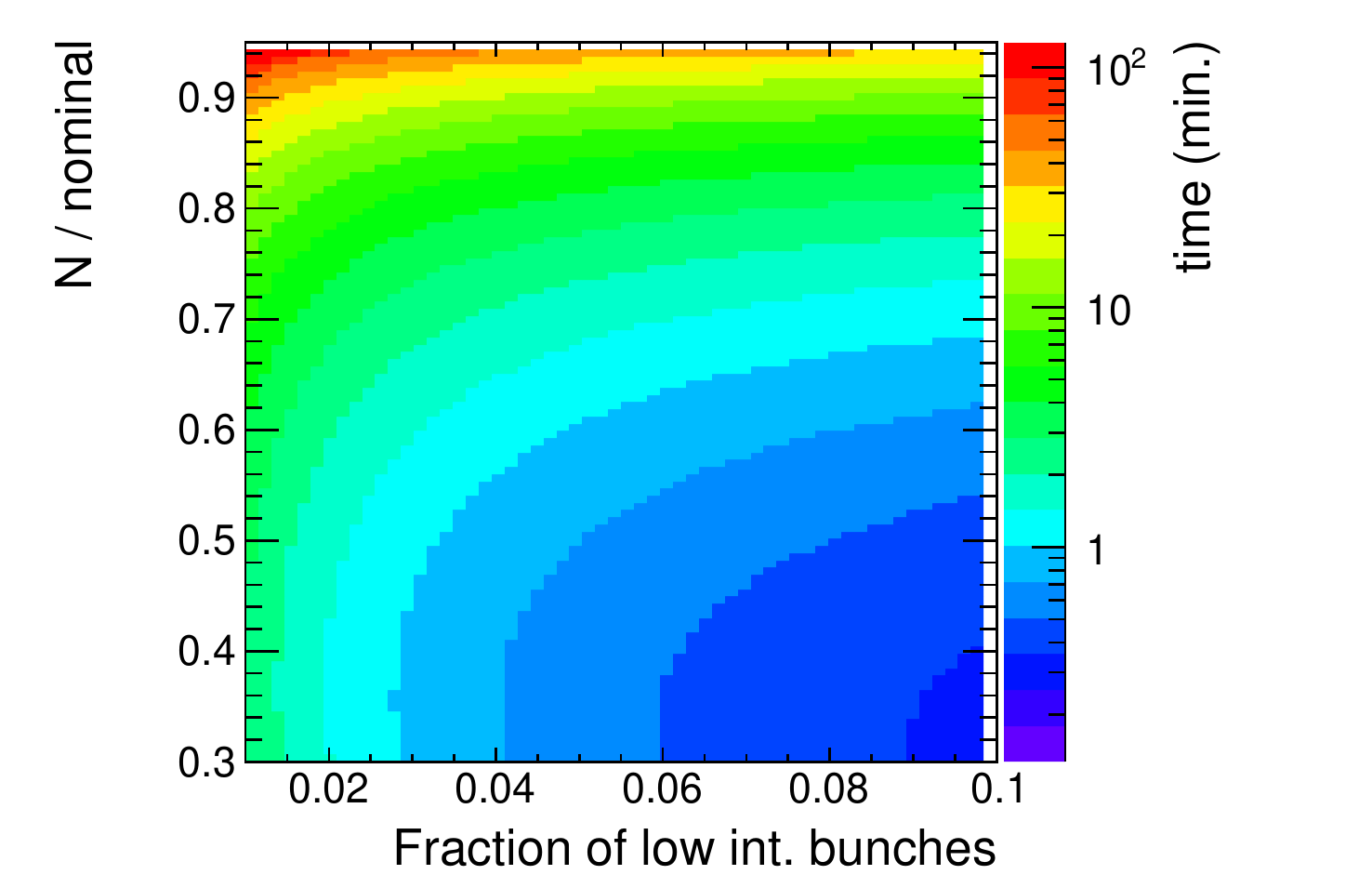}
\includegraphics[width=0.53\columnwidth]{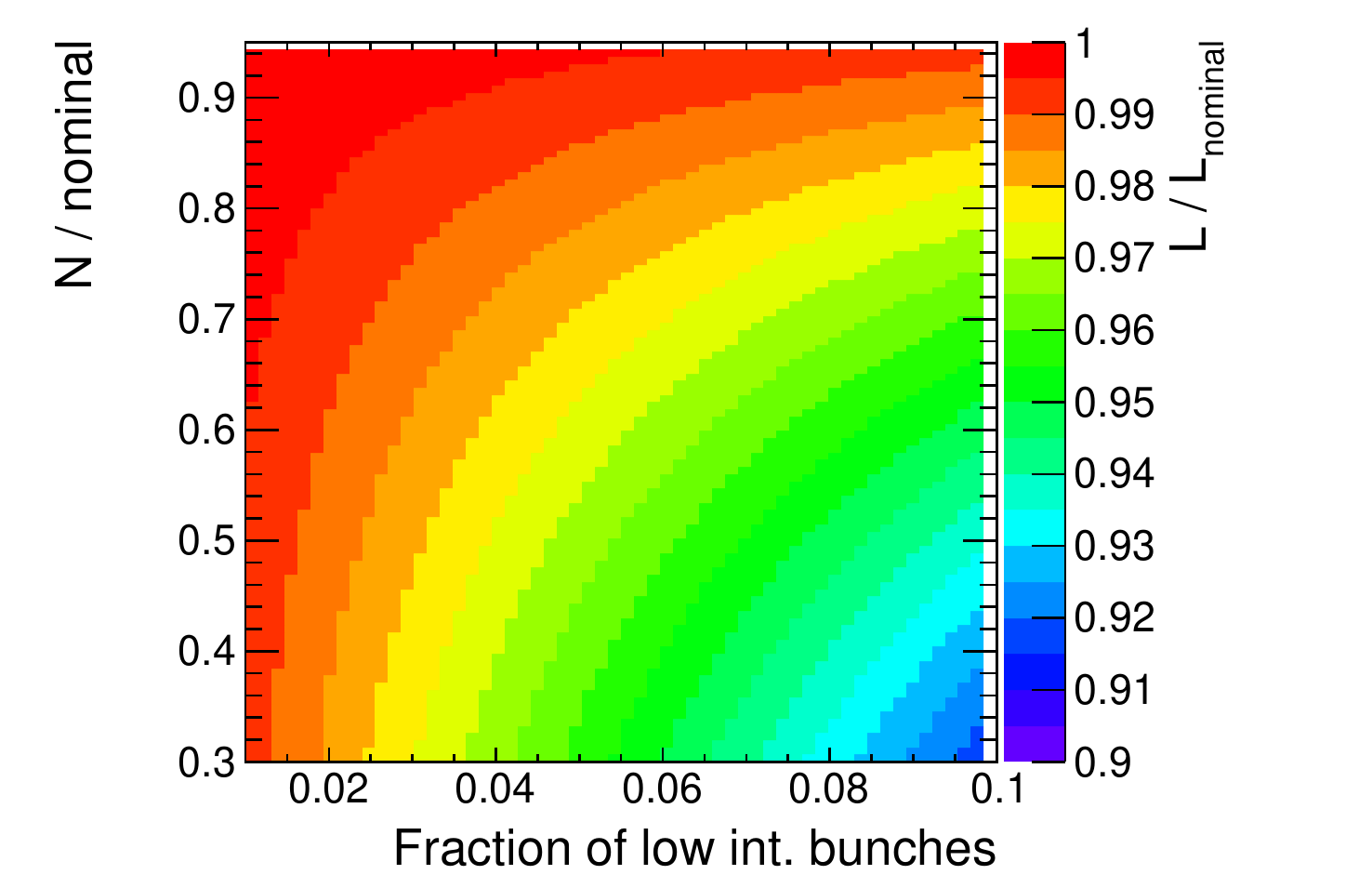}
\end{tabular}
\caption{\small Left: Time needed to measure {\tt Acol}$_{\rm{beam}}$ with a relative uncertainty of $5 \%$ in a setup with a given fraction (shown on the $x$ axis) of pilot bunches with a lower than nominal intensity (shown on the $y$ axis). The values of {\tt Acol} predicted for leading-order Bhabha events have been used. Right: Fraction of the nominal luminosity that is kept in such a setup.
}
\label{fig:asym_pilot}
\end{center}
\end{figure}

Another possibility could be to run with a setup in which a small fraction of ``pilot'' colliding bunches would have a lower intensity. The two components of {\tt Acol} could be disentangled via two measurements, one with the nominal bunches, the other with the pilot bunches. The intensities of the $\rm e^-$ and $\rm e^+$ pilot bunches are taken to be equal, and the bunch length scales as the square root of this intensity~\cite{Zimmermann:2014mza}. 
The left panel of Fig.~\ref{fig:asym_pilot} shows the time that would be needed to measure the beam-induced component of {\tt Acol} with a relative uncertainty of $5 \%$, as a function of the fraction of pilot bunches and of their intensity. Having a very low intensity for the pilot bunches  provides a larger lever arm, but because of the low luminosity of these bunches, such working points are not optimal. The plot shows that it would be possible to measure {\tt Acol}$_{\rm{beam}}$ with the required precision on the time scale of a  few minutes, 
for a limited loss in luminosity (shown in the right panel of Fig.~\ref{fig:asym_pilot}). For instance, with $2\%$ of bunches at $85 \%$ of the nominal intensity, {\tt Acol}$_{\rm{beam}}$ could be determined within $10$ minutes, for a loss in luminosity of less than $0.5 \%$.

\section{Conclusions}

Electromagnetic effects caused by the very large charge densities of the FCC-ee beam bunches affect the colliding particles in several related ways. The final state electrons and positrons from small angle Bhabha scattering are focused by the electromagnetic fields of the counter-rotating bunches leading to a sizeable bias of the luminometer acceptance, that must be corrected for in order to reach the desired precision on the luminosity measurement. Several sets of measurements can be used to control this bias.
The crossing angle increase from its nominal value that is induced by beam-beam effects, which is measured using the central detector, and an acollinearity variable that is measured using the luminometer system,  provide a determination of the correction factor to be applied to the luminosity.  For the latter measurement, the effects induced by the beams can be disentangled from those caused by a misalignment of the luminometer system.  Each of the proposed measurements allows a determination of the bias that ensures a residual uncertainty  on the absolute luminosity smaller than $10^{-4}$. In practice, they are likely to be combined, possibly with other measurements which, as the ones proposed here, are sensitive to the beam-beam interactions.

An energy scan around the $\rm Z$ peak, allowing a detailed study of the $\rm Z$ line shape, is a crucial part of the FCC-ee physics programme. A precision of a few tens of keV on the width of the $\rm Z$ boson could be reached provided that the relative luminosity of the datasets taken at the different energies is known with a precision of ${\cal{O}} ( 10^{-5} ) $.
The ramp-up period of the machine allows the luminosity bias to be determined with a statistical uncertainty of $1-2 \%$ for a given fill, which corresponds to a residual uncertainty of $2-3 \cdot 10^{-5}$ on the luminosity. Summing up over the fills taken at each energy point reduces this statistical error to below $10^{-5}$. The systematic component of the uncertainty of the luminosity bias, that arises from the uncertainty of the dependence of the acollinearity variable on the longitudinal bunch length, is largely correlated from point to point. Consequently, the beam-beam effects are not expected to contribute significantly to the uncertainty of the relative normalisation.

In the context of the studies reported here, it has been realised that, despite the smaller charge density of the bunches, the focusing of the leptons emerging from a Bhabha interaction induced by the opposite charge beam  was already impacting the luminosity measurement at the LEP collider. This is quantified in a separate paper~\cite{Voutsinas:2019hwu} in which possible corrections of this effect in the absence of an effective collision crossing angle are also outlined.

\subsection*{Acknowledgments}
We are grateful to Daniel Schulte, Helmut Burkhardt, Nicola Bacchetta, Konrad Elsener, Dima El Khechen, Mike Koratzinos, Katsunobu Oide, and Dmitry Shatilov for  very useful discussions, suggestions and input that they have brought into this work.


\bibliographystyle{jhep}
\bibliography{biblio}

\end{document}